\pdfoutput=1
\documentclass[11pt]{article}

\usepackage[preprint]{acl}

\usepackage{times}
\usepackage{latexsym}

\usepackage[T1]{fontenc}
\usepackage[utf8]{inputenc}

\usepackage{microtype}

\usepackage{inconsolata}
\usepackage[moderate]{savetrees}

\usepackage{graphicx}
\usepackage{enumitem}
\usepackage{hyperref}
\usepackage{amsmath}
\usepackage{amssymb}

\usepackage{booktabs} % 提供 \toprule, \midrule, \bottomrule
\usepackage{tabularx} % 代替 tabu 以确保兼容性
\usepackage{lscape} % 如果需要旋转表格
\usepackage{caption} % 处理表格标题
\usepackage{array} % 额外的表格格式控制
\usepackage{multirow}

\usepackage{lipsum}
\usepackage{tcolorbox}
\tcbuselibrary{breakable}

\definecolor{deepblue}{RGB}{0,51,102}

\newenvironment{dialogue}[1]{
    \tcolorbox[
        breakable=true,
        sharp corners=downhill,
        title=#1,
        width=\linewidth,
        colback=white,
        colframe=deepblue,
        colbacktitle=deepblue,
        coltitle=white
    ]
}{
    \endtcolorbox
}

\title{The Hidden Strength of Disagreement: Unraveling the Consensus-Diversity Tradeoff in Adaptive Multi-Agent Systems}

\author{
    Zengqing Wu \quad \quad \quad Takayuki Ito \\
    Graduate School of Informatics, Kyoto University \\
    \texttt{wuzengqing@outlook.com, ito@i.kyoto-u.ac.jp}
}

\begin{document}
\maketitle
\begin{abstract}
Consensus formation is pivotal in multi-agent systems (MAS), balancing collective coherence with individual diversity. Conventional LLM-based MAS primarily rely on explicit coordination, e.g., prompts or voting, risking premature homogenization. We argue that implicit consensus, where agents exchange information yet independently form decisions via in-context learning, can be more effective in dynamic environments that require long-horizon adaptability. By retaining partial diversity, systems can better explore novel strategies and cope with external shocks. We formalize a consensus-diversity tradeoff, showing conditions where implicit methods outperform explicit ones. Experiments on three scenarios -- Dynamic Disaster Response, Information Spread and Manipulation, and Dynamic Public-Goods Provision -- confirm partial deviation from group norms boosts exploration, robustness, and performance. We highlight emergent coordination via in-context learning, underscoring the value of preserving diversity for resilient decision-making. 
\end{abstract}

\def\thefootnote{}\footnotetext{Our source code is available at \url{https://github.com/wuzengqing001225/ConsensusDiversityTradeoffMAS}.}

\section{Introduction}

Multi-agent systems (MAS) have long studied how autonomous agents coordinate to achieve shared objectives in domains such as disaster response, resource allocation, information management, and task solving~\cite{chen2023multi, curșeu2017stakeholder, hongmetagpt, qian2024chatdev}. The recent advent of large language models (LLMs) as general-purpose agents~\cite{li2023camel, wu2023smart, xing2024designing} presents novel opportunities for MAS: LLM agents can dynamically exchange information, interpret instructions, and reason in natural language. This flexible communication paradigm potentially enables more \emph{human-like} approaches to consensus formation, diverging from rigid algorithms in conventional distributed systems.

\begin{figure}[t]
    \centering
    \includegraphics[width=1\linewidth]{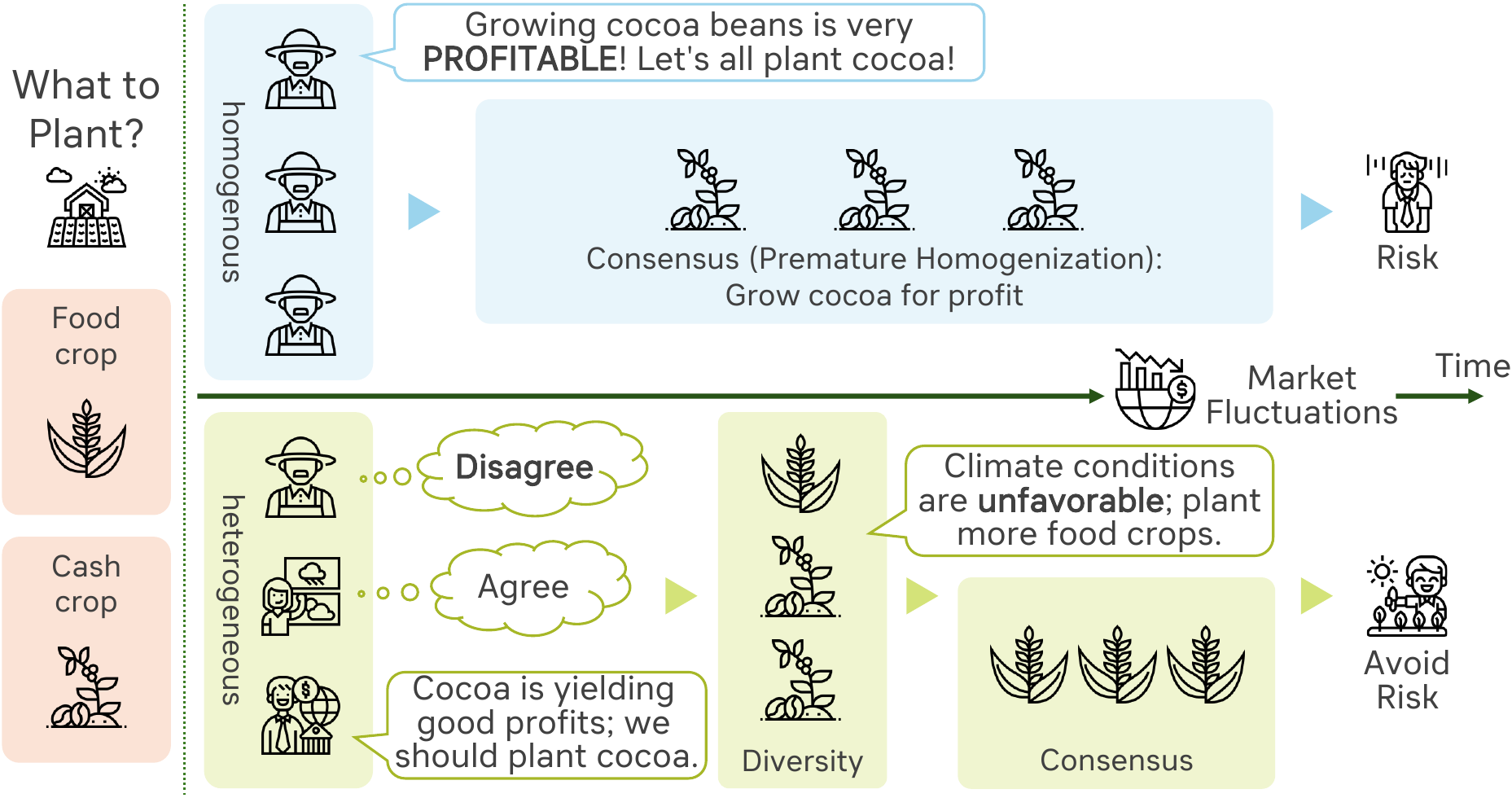}
    \vspace{-1.5em}
    \caption{An illustration of the concept of consensus-diversity tradeoff, using crop selection to show how concentrated opinions limit adaptation and lead to path dependence, which is a common real-world issue.}
    \vspace{-1.5em}
    \label{fig:illustration}
\end{figure}

However, an important challenge emerges: while strong \emph{explicit consensus} (e.g., centralized voting or forced agreement prompts) can unify the system (e.g., multi-agent debates~\cite{chanchateval}), it risks extinguishing critical diversity in agent opinions, limiting exploration and adaptability. Drawing on social science perspectives -- particularly the notion of \emph{limited collective common sense}~\cite{whiting2024framework} which suggests that collective agreement is often context-dependent and rarely complete, and that individual "common sense" can be highly idiosyncratic -- we observe that human collaboration seldom relies on achieving full, universal consensus. Instead, this theory, along with empirical observations~\cite{chen-etal-2024-reconcile, dippel2024eliminating, duan2024enhancing, shang2019resilient}, indicates that maintaining a degree of viewpoint diversity, fostering partial alignment rather than enforcing homogeneity, and tolerating individual deviations often yields more robust and adaptive group outcomes. This is especially true in uncertain or dynamically changing environments where rigid consensus can lead to premature convergence on suboptimal solutions.

Motivated by these insights from social science, which highlight that universal consensus is often unsustainable and that partial disagreement can foster adaptability, this paper proposes a \textbf{dynamic consensus-diversity tradeoff} that addresses the tension between shared understanding and autonomy in LLM-based MAS. Our key hypothesis is that \emph{implicit consensus}, in which agents discuss but act based on their own subjective interpretations (via in-context learning and potentially influenced by unique roles), can outperform explicit consensus in tasks with high environmental volatility and the need for persistent exploration. We anticipate an inverted-U relationship between inter-agent diversity and performance, where moderate diversity leads to optimal outcomes. By allowing each agent's internal chain-of-thought to incorporate external signals yet still maintain independence, the group collectively retains a broader search space of strategies, mitigating groupthink risks~\cite{shang2019resilient} and providing higher resiliency to unexpected shifts. Figure~\ref{fig:illustration} demonstrates the concept of this consensus-diversity tradeoff.

\paragraph{Contributions.}

(1)~We present a new framework for LLM-driven multi-agent implicit consensus, defining how to quantify behavioral alignment and tolerance windows for diversity.  
(2)~We propose metrics to assess \textit{when and why} implicit consensus can outperform explicit coordination, shedding light on how moderate deviations can enhance system performance.  
(3)~We validate these ideas on three different scenarios demonstrating that an in-context, discussion-based approach leads to significantly higher robustness against shocks like black swan events and adversarial behaviors.

In contrast to prior works that focus on forced alignment, single-step voting, or preset solidified agent roles~\cite{al2024project, li2023camel}, our approach does not fine-tune the model nor rely on explicit majority rule. Instead, we exploit LLMs' innate capacity for in-context learning, enabling them to interpret repeated dialogues among agents and adapt to emergent cues~\cite{han2023guinea,li2023camel,wu-etal-2024-shall,xing2024designing}. Through controlled experiments, we reveal how partial heterogeneity in agent preferences fosters resilience, aligning with social and cognitive theories emphasizing that incomplete consensus can yield robust group decisions.

\section{Related Work}

\paragraph{Emerging Role of LLMs in Multi-Agent Systems} Recent advances have started leveraging LLMs as autonomous agents within MAS~\cite{chen-etal-2024-llmarena,islam-etal-2024-mapcoder,wang-etal-2024-rethinking-bounds}. Unlike traditional MAS with fixed protocols, LLM-based agents can dynamically communicate and coordinate via natural language, enabling more flexible collaboration. Early demonstrations show that multiple LLM agents working together can solve complex tasks beyond the capability of a single model. For example, frameworks like \texttt{CAMEL} employ two ChatGPT-based agents in complementary roles (e.g. user and assistant) to cooperatively complete tasks through iterative dialogue \cite{li2023camel}. Similarly, HuggingGPT-style approaches orchestrate multiple specialized models guided by an LLM, hinting at the potential of MAS-driven problem solving. More recently, Generative Agents have been introduced as an application of LLM-based MAS in interactive simulations of social environments \cite{gao2024large,huang2024social,park2023generative}. In this paradigm, dozens of LLM-driven agents simulated believable human behaviors and social interactions over time, demonstrating new use cases of MAS in social simulations and digital environments.

\paragraph{Collaboration and Consensus Mechanisms in LLM-Based MAS} Effective collaboration and consensus in LLM-based MAS are key research areas. Traditional game theory models of consensus (e.g., Nash equilibrium~\cite{fujita2014approach,pramanik2021consensus,ye2017distributed}) predate LLM agents. Recent works explore LLM-specific interaction patterns for collective reasoning. For instance, Du et al.~\cite{Du2023improving} utilize multi-agent debate for factual consensus, whereas our work examines the consensus-diversity tradeoff, highlighting benefits of moderate, autonomy-preserving disagreement for dynamic adaptation. Debate protocols~\cite{liu2024groupdebate,zhang-etal-2024-exploring} and voting mechanisms like \texttt{RoundTable}~\cite{Cho2024roundtable} are also studied, often for static tasks. Our research on dynamic settings complements these, suggesting that full consensus may not always be optimal for adaptability. Hierarchical roles (e.g., \texttt{MAgICoRe}~\cite{Chen2024magicore}) and self-consistency via multi-path reasoning and voting~\cite{Wang2023selfconsistency} also implicitly use ensemble opinions for consensus. A survey by Guo et al.~\cite{Guo2024survey} notes that communication, memory, and conflict resolution are vital for LLM agent coordination, with knowledge consistency being an open challenge~\cite{zhang-etal-2024-exploring,Guo2024survey}. Wang et al.~\cite{wang-etal-2024-rethinking-bounds} caution that multi-agent discussion benefits depend on interaction design and task difficulty.

\paragraph{Challenges and Advances in Knowledge Integration for LLM-Based MAS} LLM-based MAS builds upon prior work in multi-agent AI, including emergent communication in RL agents~\cite{Foerster2016communicate}, which foreshadowed LLMs' natural language cooperation. LLMs offer implicit coordination through pre-trained knowledge but require careful alignment to manage error propagation and bias~\cite{wang-etal-2024-rethinking-bounds,Guo2024survey}. Their nuanced interpretation capabilities enhance human-like negotiation beyond traditional MAS.

\paragraph{Balancing Consensus and Diversity in Group Problem-Solving} Research in social and cognitive sciences highlights the necessity of balancing consensus and diversity in group problem-solving. While consensus enables coordination, diversity fosters creativity and robustness. Probabilistic opinion dynamics studies suggest that evolving opinions dynamically aids decision-making~\cite{liu2022probabilistic}. Hong \& Page~\cite{Hong2004diversity} show that diverse agent groups can outperform homogeneous high-performers on complex problems. In LLM-based MAS, fostering diverse hypotheses before convergence enhances outcomes. Smaldino et al.~\cite{Smaldino2024diversity} emphasize "transient diversity," where delayed convergence improves problem-solving. Inspired by these findings, MAS research is designing protocols that integrate diverse reasoning while ensuring eventual coherence.

\section{Methodology}
\label{sec:methodology}

\begin{figure}[t]
    \centering
    \includegraphics[width=1\linewidth]{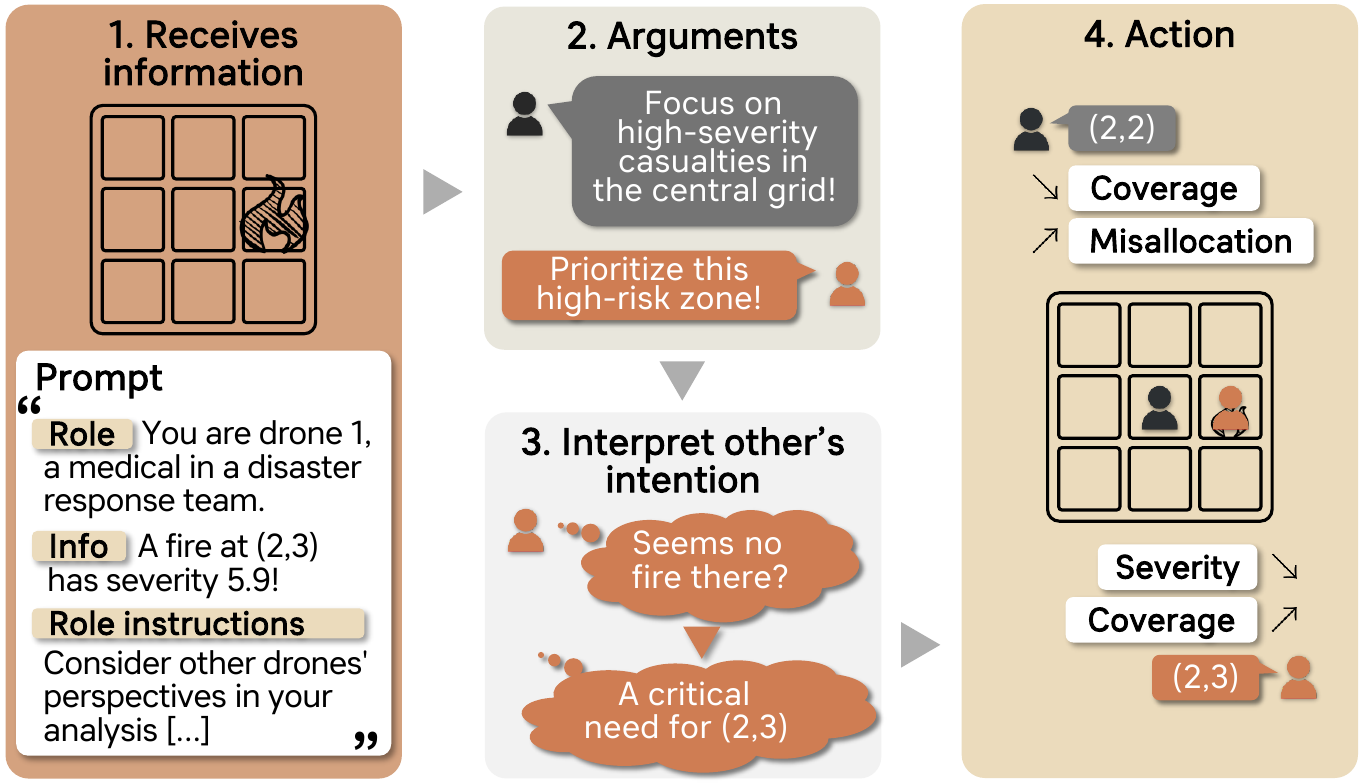}
    \vspace{-1.em}
    \caption{Conceptual workflow of multi-agent interaction. In explicit consensus, agents determine a collective action (e.g., via voting). In implicit consensus (illustrated for the disaster response scenario), agents individually interpret shared information (discussion) and then independently choose their actions, allowing for diversity.}
    \vspace{-1.em}
    \label{fig:workflow}
\end{figure}

Our framework is evaluated across three dynamic scenarios: \textbf{(1) Dynamic Disaster Response} (agents manage resources for evolving crises), \textbf{(2) Information Spread and Manipulation} (agents counter misinformation in a network), and \textbf{(3) Dynamic Public Goods Provision} (agents contribute to a public good with fluctuating requirements). These scenarios share common traits like volatile environments and the need for collaborative adaptation. The main text details the Disaster Response scenario to illustrate core concepts, while the other two (see Appendix~\ref{sec:appendix:a}) demonstrate generalizability. Figure~\ref{fig:workflow} shows the workflow of our case studies.

We motivate our approach by drawing parallels with human collaboration, which rarely achieves full consensus but often thrives on partial alignment and diverse perspectives---a concept related to "limited collective common sense"~\cite{whiting2024framework}. Our goal is to formalize and empirically investigate how controlled deviation from complete consensus can enhance adaptability and robustness in dynamic environments.

\subsection{Framework: Implicit vs. Explicit Consensus}
\label{ssec:framework}

We consider $N$ LLM agents $\{1, 2, \dots, N\}$ collaborating over discrete rounds $t \in \{1,2,\ldots\}$ to address dynamic tasks. Each round, agents typically:
\begin{enumerate}[itemsep=0pt, parsep=0pt, leftmargin=*]
    \item \emph{Observe}: Receive information about the current state of the environment and transcripts of messages from other agents from previous rounds (communication may be partial or noisy).
    \item \emph{Communicate}: Generate textual messages containing their assessments, proposals, arguments, or evidence to a shared discussion $D(t)$.
    \item \emph{Act}: Based on the available information and discussion, each agent $i$ commits to an action $a_i(t)$. The mechanism for this commitment distinguishes our two primary modes of consensus.
\end{enumerate}

\paragraph{Explicit Consensus.}
In this mode, agents are compelled to reach a unified decision. They first propose individual actions or plans, and then a collective action is determined, typically through a mechanism like majority voting:
\begin{equation}
    a_{\text{collective}}(t) = \arg\max_{a \in \mathcal{A}} \sum_{j=1}^N \mathbb{I}[v_j(t) = a],
\end{equation}
where $v_j(t)$ is the action proposed by agent $j$ and $\mathcal{A}$ is the action space. All agents then adopt this collective action: $a_i(t) = a_{\text{collective}}(t)$ for all $i$. Alternatively, forced alignment can be achieved via strong prompting (e.g., ``All agents must agree on and execute the exact same action.''). This approach ensures coordination but can suppress beneficial diversity, potentially leading to premature convergence or suboptimal performance in dynamic or complex tasks. Agents \textbf{do not} use an individual \textit{interpret} step after the collective decision is made; they simply adopt the group's choice.

\paragraph{Implicit Consensus.}
Here, agents engage in discussion but retain autonomy in their final action choices. Each agent $i$ independently \emph{interprets} the shared discussion $D(t)$ and other observations, then selects its action $a_i(t)$:
\begin{equation}
    a_i(t) \,\sim\, \textsc{LLM}_i\Bigl(\scriptstyle\text{``Given discussion } D(t)\text{, choose action''}\Bigr),
\end{equation}
The \textit{interpret} step is crucial: it allows for subjective assessment of the group discourse, enabling agents to maintain individual perspectives (potentially shaped by unique roles or information) while still being influenced by others. This subjective interpretation means that even after discussion, agents might choose different actions. This mechanism allows for a natural emergence of partial alignment and diversity.

\paragraph{Role Prompts and Specialization.}
To inject controlled diversity, agents can be assigned unique \textbf{role prompts} (e.g., in a disaster response scenario: ``You are a medical drone, prioritize high-casualty zones''; ``You are a logistics drone, prioritize delivering supplies''). These textual descriptions bias agents' decision-making within a shared action space, promoting varied perspectives and strategies without prescribing mutually exclusive actions. All agents, regardless of role, choose from the same fundamental action pool (e.g., grid coordinates in Scenario 1). Specialization influences preferences, not the set of possible actions.

\subsection{The Dynamic Consensus-Diversity Model}
\label{ssec:dynamic_model}
We aim to quantify the system's collective behavior and the diversity of actions.

\paragraph{System's Action Distribution and Mean Action.}
The system's collective state at round $t$ is captured by the empirical distribution of agent actions $C(t)$:
\begin{equation}
    C(t) = \frac{1}{N}\sum_{i=1}^N \delta(a_i(t)),
\end{equation}
where $\delta$ is the Dirac delta function. For \texttt{discrete (categorical) actions} (e.g., grid coordinates in Scenario 1, node sets in Scenario 2), $\delta(a_i(t))$ acts as an indicator function. $C(t)$ then represents the frequency of each action. For example, if 5 agents choose actions [A, A, A, B, B], then $C(t) = \{\text{A: } 0.6, \text{B: } 0.4\}$. For \texttt{continuous actions} (e.g., contribution levels in Scenario 3), $\delta(a_i(t))$ can be generalized to a probability density function (e.g., a narrow Gaussian kernel centered at $a_i(t)$) to model the distribution of actions.

The mean action $\mu(t)$ is the central tendency of $C(t)$. For \texttt{discrete actions}, $\mu(t)$ is the mode (i.e., most frequent action). In the example above, $\mu(t) = A$. For \texttt{continuous actions}, $\mu(t) = \mathbb{E}_{a \sim C(t)}[a]$ is the arithmetic mean.

\paragraph{Measuring Action Deviation.}
The deviation of agent $i$'s action from the mean action is $d_i(t) = \text{distance}(a_i(t), \mu(t))$. The specific distance metric depends on the action space:

\begin{itemize}[itemsep=0pt, parsep=0pt, leftmargin=*]
    \item Scenario 1 (Disaster Response, grid coordinates): Manhattan distance, $d_i(t) = |x_i - x_\mu| + |y_i - y_\mu|$.
    \item Scenario 2 (Information Spread, sets of nodes): Jaccard distance, $d_i(t) = 1 - \frac{|S_i \cap S_\mu|}{|S_i \cup S_\mu|}$, capturing overlap in selected node sets.
    \item Scenario 3 (Public Goods Provision, continuous contribution): Normalized absolute difference, $d_i(t) = \frac{|c_i(t) - \mu(t)|}{C_{\text{max}}}$.
\end{itemize}
The average deviation across all agents is $\bar{d}(t) = \frac{1}{N}\sum_{i=1}^N d_i(t)$. This $\bar{d}(t)$ quantifies the degree of disagreement or diversity in actions at round $t$. Even with specialization, actions are often correlated by task dynamics (e.g., disasters are localized), preventing maximal disagreement.

\paragraph{The Inverted-U Hypothesis.}
We hypothesize an \emph{inverted-U relationship} between average deviation $\bar{d}(t)$ and system performance. Performance is expected to be low with very low diversity (premature consensus, $\bar{d}(t) \approx 0$) and also with very high diversity (loss of coordination). Optimal performance is anticipated at a moderate level of $\bar{d}(t)$, where the system balances exploration and exploitation. This aligns with findings in social sciences that suggest incomplete consensus can foster robustness~\cite{Hong2004diversity, Smaldino2024diversity}.

\subsection{Other Concerns on Coordination}
\label{ssec:stability_coordination}

A concern with implicit consensus is whether the system can achieve stable and effective coordination.

\paragraph{Emergent Coordination via In-Context Learning.}
LLM agents adapt their reasoning and actions based on natural language dialogue. The conversation itself, influenced by role prompts and environmental cues, becomes the medium for coordination. This NLP-centric mechanism is distinct from traditional MAS where coordination protocols are often explicitly programmed.

\paragraph{Theoretical Cross-Validation.}
To better understand the fundamental dynamics of consensus and diversity, particularly the difference between structured, role-based diversity and purely random deviations, we employ a simplified random-iteration theoretical model. This model, detailed in Appendix~\ref{app:theoretical_model}, helps establish a baseline by showing that unstructured noise typically degrades performance. This contrasts with our main experimental findings where meaningful, LLM-driven diversity enhances adaptability. The model explores conditions for convergence and the impact of random shocks, providing context for the empirically observed benefits of purposeful heterogeneity.

\paragraph{Scalability Considerations.}
While our experiments use up to $N=100$ agents with direct discussion, larger systems might require hierarchical or parallel communication structures to manage overhead. Such extensions are beyond this study's current empirical scope.

\paragraph{Agent Formalism in Scenario 1 (Disaster Response).}
To make the formalism concrete for Scenario 1: each of the $N$ LLM agents controls one of $N$ drones (a 1:1 mapping). The action $a_i(t)$ for LLM agent $i$ is the grid coordinate its drone will move to in round $t$.
\begin{itemize}[itemsep=0pt, parsep=0pt, leftmargin=*]
    \item In \textbf{explicit consensus}, all LLM agents propose a target coordinate; these proposals are aggregated (e.g., by majority vote), and the single winning coordinate becomes the action $a_{\text{collective}}(t)$ for \emph{all} $N$ drones in that round.
    \item In \textbf{implicit consensus}, each LLM agent, after reviewing the shared discussion $D(t)$ and considering its role prompt, independently decides the target coordinate $a_i(t)$ for its own drone. This can result in drones moving to different locations based on their controlling LLM's interpretation.
\end{itemize}

\section{Experimental Setup}
\label{sec:experiments}

We design three dynamic scenarios to evaluate our research questions:
\begin{enumerate}[itemsep=1pt, parsep=1pt, leftmargin=*]
    \item \textbf{Q1}: Does implicit consensus outperform explicit coordination in volatile or adversarial conditions, and how do different LLM models affect this?
    \item \textbf{Q2}: Under what conditions do moderate deviations (diversity) improve system robustness, aligning with the inverted-U hypothesis?
    \item \textbf{Q3}: How do LLM agents' in-context chain-of-thought updates and dialogue reflect an evolving consensus and coordination, particularly in the implicit setting?
\end{enumerate}

\subsection{Scenario Overview and Key Mechanics}
\label{ssec:scenario_overview}
We utilize three distinct scenarios for our experiments. The main text will primarily focus on Scenario 1 for detailed illustration. Full descriptions, including environmental dynamics, agent roles, and specific parameters for all three scenarios (Dynamic Disaster Response, Information Spread and Manipulation, and Dynamic Public Goods Provision) are provided in Appendix~\ref{sec:appendix:a}.

\paragraph{Scenario 1: Dynamic Disaster Response.}
Autonomous LLM-piloted drones operate on a $10\times10$ grid to manage resources (e.g., firefighting, medical aid) for disasters of varying severity. Disaster locations and severities can change unpredictably each round, communicated through textual environmental reports which may sometimes be contradictory or incomplete. Agents decide on grid cells to target.

\subsection{Core Metrics}
\label{ssec:core_metrics}
To evaluate performance, we use scenario-specific key metrics. Detailed definitions for all metrics across all scenarios, alongside hyperparameter settings, are available in Appendix~\ref{sec:appendix:b}. Metrics in Scenario 1 (Dynamic Disaster Response):

\begin{itemize}[itemsep=0pt,parsep=0pt,leftmargin=*]
    \item \textbf{Coverage Rate (CR)}: Percentage of active disaster cells correctly attended by at least one agent. Higher is better.
    \item \textbf{Misallocation Penalty (MP)}: Penalty incurred for assigning agents to non-disaster cells or over-allocating to low-severity cells when high-severity cells are unattended. Lower is better.
    \item \textbf{Response Delay (RD)}: Average number of rounds taken to attend to a new high-severity disaster after its appearance. Lower is better.
\end{itemize}

\subsection{Experimental Design}
\label{ssec:exp_design}
Each scenario is run for $T=20$ to $30$ rounds. In every round, agents first \emph{observe} environmental states and textual updates (which can be partial or contradictory). They then \emph{communicate} by engaging in one or two turns of dialogue, generating textual messages based on their roles and observations. Finally, they \emph{act}: actions are determined either via forced voting for \textbf{explicit consensus}, or through individual decisions after discussion for \textbf{implicit consensus}. After actions are taken, we compute agent deviations $d_i(t)$ and aggregate performance metrics.

We vary the following factors:
\begin{itemize}[itemsep=1pt, parsep=1pt, leftmargin=*]
    \item \textbf{Diversity Level}: \textit{Low} (all agents share identical role prompts), \textit{Medium} (2--3 distinct, cooperative roles), \textit{High} (more distinct roles, potentially with some conflicting individual goals compatible with overall task).
    \item \textbf{Volatility Level}: \textit{Low} (infrequent environmental changes), \textit{Moderate} (periodic changes), \textit{High} (frequent shocks or adversarial actions).
    \item \textbf{LLM Variants}: Experiments utilize a range of models as detailed in Section~\ref{sec:results}. All agents within a single experimental run use the same base LLM (homogeneous teams), unless specified as a mixed-model setup.
\end{itemize}
We perform 5 runs for each experimental setting (e.g., Diversity:Medium + Volatility:High + Model:GPT-4o) to enhance reproducibility.

\begin{figure}
    \centering
    \includegraphics[width=1\linewidth]{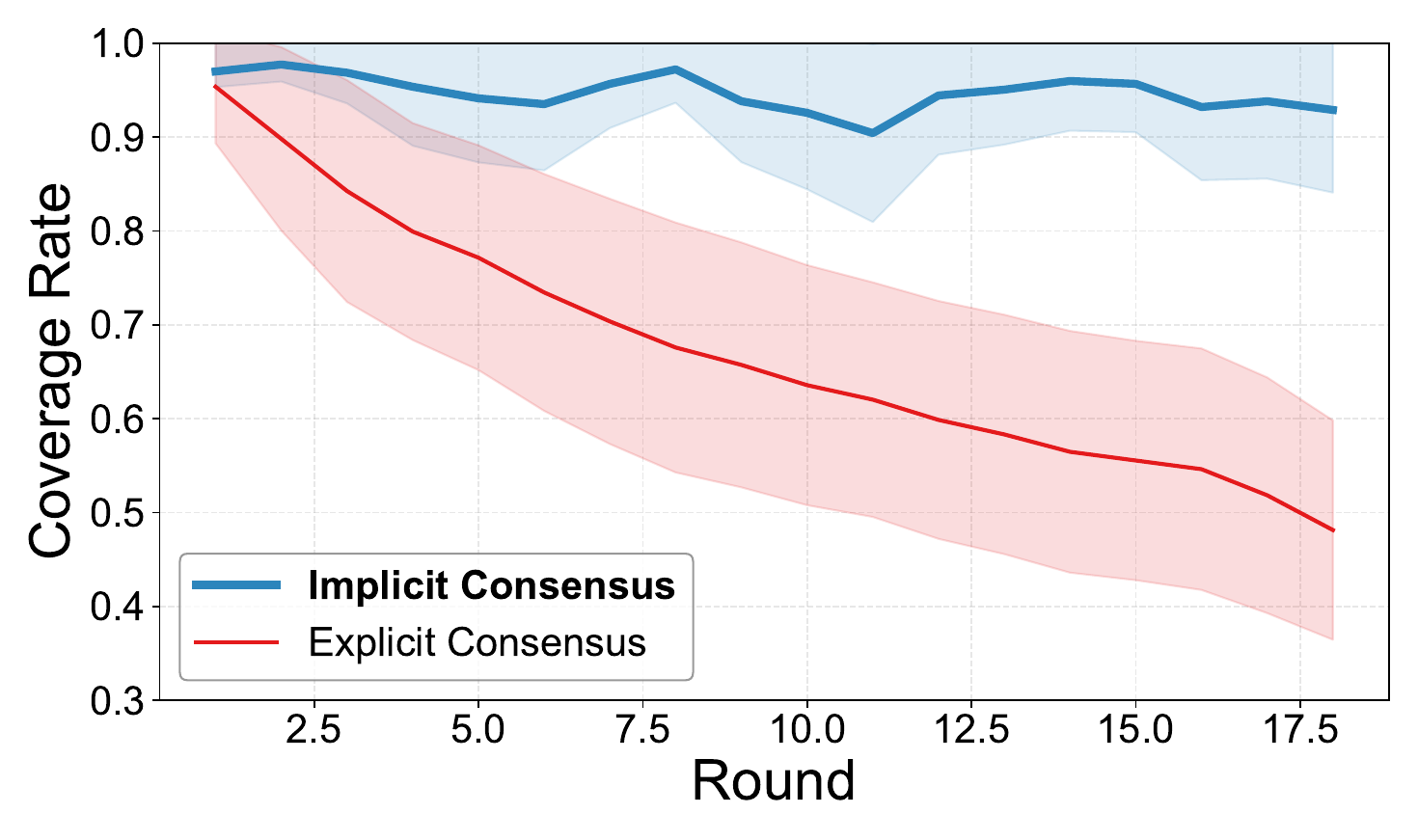}
    \vspace{-1.5em}
    \caption{Performance Comparison between Implicit Consensus and Explicit Consensus.}
    \label{fig:timeseries-main}
    \vspace{-0.5em}
\end{figure}

\begin{figure}
    \centering
    \includegraphics[width=1\linewidth]{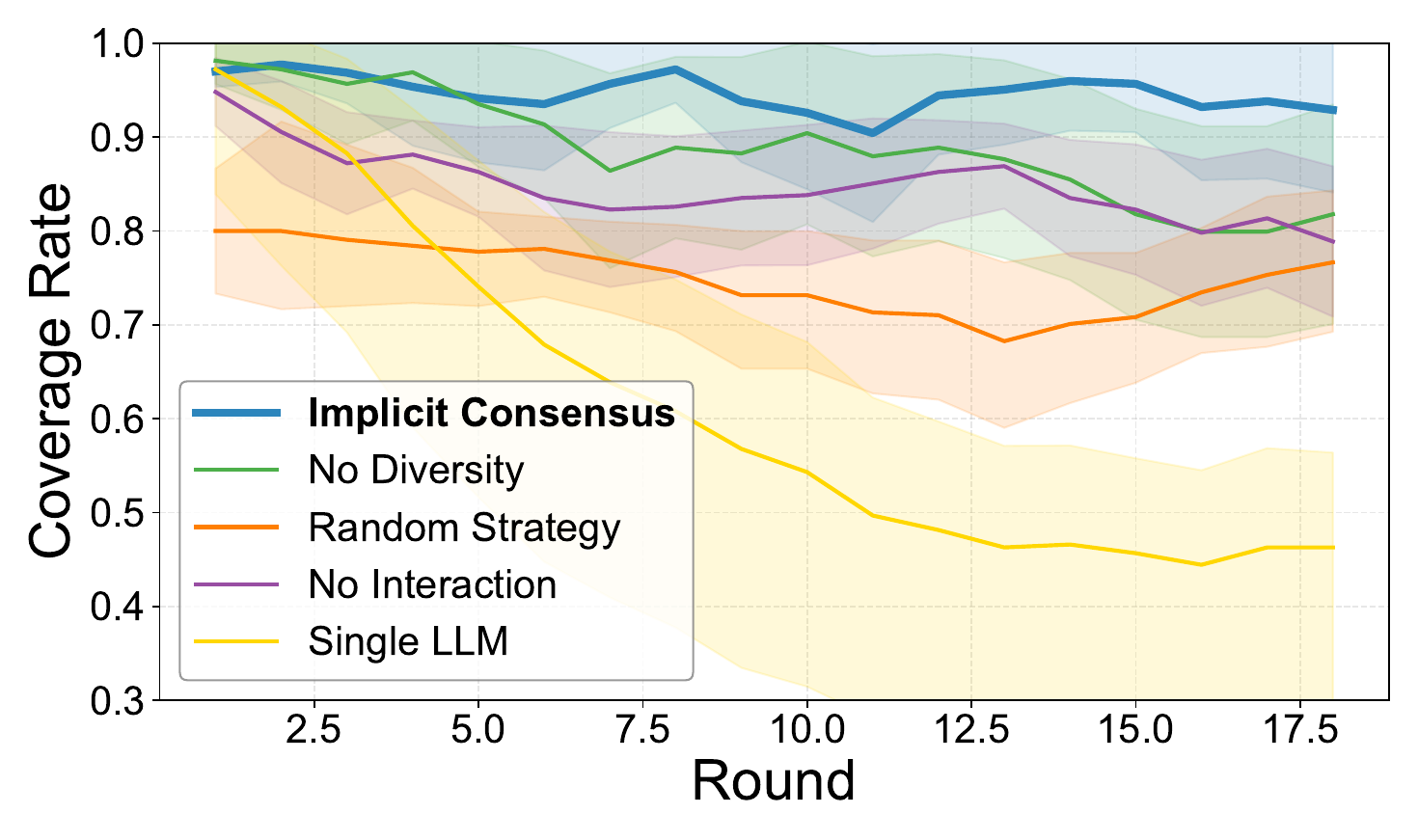}
    \vspace{-1.5em}
    \caption{Performance Comparison between Implicit Consensus and other baselines.}
    \vspace{-1.5em}
    \label{fig:timeseries-ablation}
\end{figure}

\subsection{Baselines and Comparison Protocols}
\label{sssec:baselines_protocols}

To evaluate the effectiveness of our proposed implicit consensus mechanism with role-based diversity, we compare it against several baselines and alternative configurations:

\begin{itemize}[itemsep=0pt, parsep=0pt, leftmargin=*]
    \item \textbf{Explicit Consensus (EC)}: Agents are forced to agree on a single action, typically via majority voting on proposed actions or through strong instructional prompts mandating uniformity. This serves as a primary comparison for implicit consensus.
    \item \textbf{Single-LLM}: A single LLM agent performs the task alone. This baseline helps quantify the benefits of multi-agent collaboration.
    \item \textbf{No-Interaction}: Multiple agents act independently without any communication or coordination. This highlights the value of inter-agent discussion.
    \item \textbf{Random Strategy}: Agents choose actions randomly from the available action space. This provides a lower bound on performance.
    \item \textbf{No-Diversity (Homogeneous Roles)}: All agents in the implicit consensus setting operate with identical role prompts. This ablation helps isolate the impact of role-induced diversity.
\end{itemize}

These baselines allow us to systematically assess the contributions of multi-agent interaction, the consensus mechanism (implicit vs. explicit), and the presence of agent diversity. We also analyze performance scaling with the number of agents ($N=3$ to $N=100$). Agent actions and textual messages are collected to investigate how in-context learning and linguistic cues facilitate emergent coordination and adaptation (RQ3).

\section{Results and Analysis}
\label{sec:results}

\subsection{Overall Performance Comparison (RQ1)}

\begin{table}
\centering
\small
\caption{Overall results of dynamic disaster response scenario: Comparison between Explicit and Implicit (Main) Consensus. Metrics include Coverage Rate (CR, higher is better), Misallocation Penalty (MP, lower is better), and Response Delay (RD, lower is better).}
\renewcommand{\arraystretch}{1.2}
\setlength{\tabcolsep}{4pt}
\resizebox{\linewidth}{!}{
\begin{tabular}{l l ccc ccc}
\toprule
\multirow{2}{*}{\textbf{Condition}} & \multirow{2}{*}{\textbf{Level}} & \multicolumn{3}{c}{\textbf{Explicit Consensus}} & \multicolumn{3}{c}{\textbf{Implicit Consensus}} \\
\cmidrule(lr){3-5} \cmidrule(lr){6-8}
& & \textbf{CR} & \textbf{MP} & \textbf{RD} & \textbf{CR} & \textbf{MP} & \textbf{RD} \\
\midrule
\textbf{Overall} & - & 0.679 & 2.847 & 1.324 & \textbf{0.952} & \textbf{0.208} & \textbf{0.222} \\
\midrule
\multirow{3}{*}{\textbf{Diversity}} 
& Low    & 0.671 & 2.958 & 1.352 & \textbf{0.918} & \textbf{0.542} & \textbf{0.436} \\
& Medium & 0.661 & 2.750 & 1.438 & \textbf{0.968} & \textbf{0.042} & \textbf{0.134} \\
& High   & 0.706 & 2.833 & 1.183 & \textbf{0.969} & \textbf{0.042} & \textbf{0.094} \\
\midrule
\multirow{3}{*}{\textbf{Volatility}} 
& Low     & 0.771 & 2.000 & 0.743 & \textbf{0.928} & \textbf{0.458} & \textbf{0.389} \\
& Moderate & 0.628 & 3.292 & 1.860 & \textbf{0.969} & \textbf{0.042} & \textbf{0.118} \\
& High    & 0.639 & 3.250 & 1.370 & \textbf{0.958} & \textbf{0.167} & \textbf{0.159} \\
\bottomrule
\end{tabular}
}
\vspace{-1em}
\label{tab:consensus}
\end{table}

\paragraph{Implicit vs. Explicit Consensus.}
In the disaster response scenario (Table~\ref{tab:consensus}), implicit consensus (IC) consistently outperforms explicit consensus (EC). IC achieves higher average coverage rates (CR), especially under moderate to high volatility (e.g., CR > 0.95 for IC vs. < 0.65 for EC in high volatility). EC teams tend to over-commit to single zones and adapt slowly to new disasters. In contrast, IC teams, through ongoing discussion, allow individual agents to deviate and address overlooked or emergent critical locations. Consequently, IC also shows significantly lower misallocation penalties (MP), as its inherent diversity promotes broader coverage and self-correction. Furthermore, IC demonstrates faster response times (lower mean RD) to new high-severity disasters, as agents in IC may investigate uncertain reports, accelerating detection even without full certainty. Figure~\ref{fig:timeseries-main} illustrates the performance difference over time.

\paragraph{Comparison with Other Baselines.}
Figure~\ref{fig:timeseries-ablation} extends the comparison, showing IC also surpasses other baselines. Notably,~\emph{No Diversity} systems (identical prompts) suffer from rigid decision-making, missing localized exploration opportunities.~\emph{No Interaction} leads to uncoordinated actions and high MP. \emph{Random Strategy} fails to adapt to real-time changes effectively. The~\emph{Single LLM} baseline, lacking both diversity and parallel processing, performs worst. These results confirm IC's superior adaptability in dynamic disaster conditions (affirming RQ1). Agent scale did not significantly impact results in the current setup.

\begin{table*}[t]
\centering
\small
\caption{Ablation Study: Comparing different configurations, including Main, No Diversity, No Interaction, Random Strategy, and Single LLM.}
\renewcommand{\arraystretch}{1.2}
\setlength{\tabcolsep}{4pt}
\resizebox{\linewidth}{!}{
\begin{tabular}{l l ccc ccc ccc ccc ccc}
\toprule
\multirow{2}{*}{\textbf{Condition}} & \multirow{2}{*}{\textbf{Level}} & \multicolumn{3}{c}{\textbf{Implicit Consensus}} & \multicolumn{3}{c}{\textbf{No Diversity}} & \multicolumn{3}{c}{\textbf{No Interaction}} & \multicolumn{3}{c}{\textbf{Random Strategy}} & \multicolumn{3}{c}{\textbf{Single LLM}} \\
\cmidrule(lr){3-5} \cmidrule(lr){6-8} \cmidrule(lr){9-11} \cmidrule(lr){12-14} \cmidrule(lr){15-17}
& & \textbf{CR} & \textbf{MP} & \textbf{RD} & \textbf{CR} & \textbf{MP} & \textbf{RD} & \textbf{CR} & \textbf{MP} & \textbf{RD} & \textbf{CR} & \textbf{MP} & \textbf{RD} & \textbf{CR} & \textbf{MP} & \textbf{RD} \\
\midrule
\textbf{Overall} & - & \textbf{0.952} & \textbf{0.167} & 0.222 & 0.892 & 0.750 & 0.316 & 0.843 & 0.278 & 0.246 & 0.754 & 0.208 & \textbf{0.173} & 0.628 & - & 1.619 \\
\midrule
\multirow{3}{*}{\textbf{Diversity}} 
& Low    & \textbf{0.918} & \textbf{0.458} & 0.436 & 0.906 & 0.667 & \textbf{0.227} & 0.782 & 0.750 & 0.553 & 0.714 & 0.542 & 0.330 & 0.729 & - & 1.400 \\
& Medium & \textbf{0.968} & \textbf{0.000} & 0.134 & 0.917 & 0.625 & 0.296 & 0.878 & \textbf{0.000} & 0.088 & 0.771 & 0.042 & \textbf{0.068} & 0.524 & - & 2.089 \\
& High   & \textbf{0.969} & \textbf{0.042} & 0.094 & 0.853 & 0.958 & 0.426 & 0.868 & 0.083 & 0.095 & 0.776 & \textbf{0.042} & \textbf{0.120} & 0.632 & - & 1.368 \\
\midrule
\multirow{3}{*}{\textbf{Volatility}} 
& Low     & 0.928 & \textbf{0.246} & 0.389 & \textbf{0.942} & 0.375 & 0.261 & 0.858 & 0.250 & \textbf{0.207} & 0.747 & 0.458 & 0.259 & 0.592 & - & 2.171 \\
& Moderate & \textbf{0.969} & 0.333 & \textbf{0.118} & 0.865 & 0.958 & 0.414 & 0.839 & 0.250 & 0.210 & 0.746 & \textbf{0.000} & 0.189 & 0.622 & - & 1.459 \\
& High    & \textbf{0.958} & \textbf{0.000} & 0.159 & 0.868 & 0.917 & 0.273 & 0.831 & \textbf{0.333} & 0.320 & 0.768 & 0.167 & \textbf{0.070} & 0.671 & - & 1.227 \\
\bottomrule
\end{tabular}
}
\vspace{-4pt}
\label{tab:ablation}
\end{table*}

\begin{figure}
    \centering
    \includegraphics[width=1\linewidth]{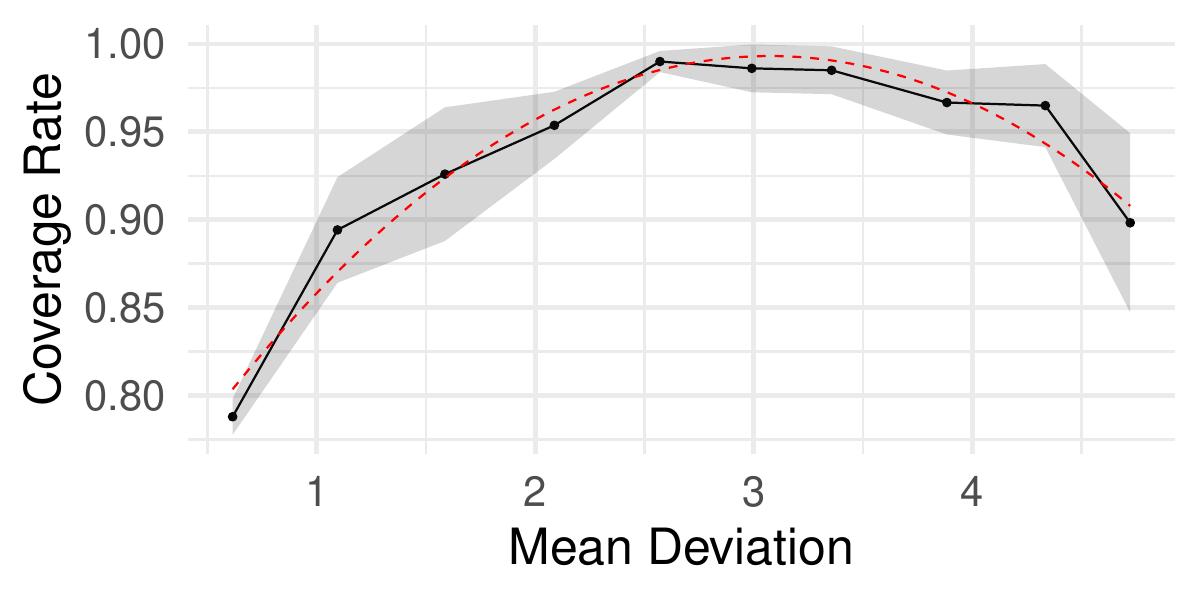}
    \vspace{-1.5em}
    \caption{Deviation-Performance Correlation Plot: Validating the Inverted U-Shaped Hypothesis.}
    \vspace{-1.5em}
    \label{fig:deviation}
\end{figure}

\subsection{Deviation-Performance Correlation (RQ2)}
\label{subsec:deviation}

While higher coverage rates and lower penalties favor IC over EC, a crucial question is \emph{why}. The \emph{dynamic consensus-diversity tradeoff} suggests that moderate agent-level deviations encourage exploration and rapid environmental adaptation. To verify this, we measure each agent's deviation from the group mean, $d_i(t)$, and then plot the average $\bar{d}(t)$ against performance metrics.

\paragraph{Inverted-U Pattern.} In Figure~\ref{fig:deviation}, we observe a clear inverted-U relationship between mean deviation (\(\bar{d}\)) and coverage rate as the performance in scenario 1. When $\bar{d} \approx 0$ (full uniformity), coverage is suboptimal (under $0.80$ on average), because the system becomes over-committed to a single or obvious priority and overlooks secondary crises. In contrast, when $\bar{d}$ is extremely large (e.g., agents rarely agree on any zone), coverage also dips, reflecting disorganized duplication or spread-thin allocations. The peak of coverage near \(\bar{d}\approx 2 \) to \(3.5\) exemplifies how partial disagreement fosters just enough diversity to handle multiple emergent zones simultaneously while still maintaining broad consensus on urgent tasks.

\paragraph{Role of Diversity Levels.} Table~\ref{tab:consensus} and the ablation in Table~\ref{tab:ablation} break down performance for \emph{low}, \emph{medium}, and \emph{high} diversity. We see that:

\begin{itemize} [itemsep=0pt, parsep=0pt, leftmargin=*]
    \item \textbf{Medium/High Diversity} tends to yield the highest coverage rates (often exceeding 0.95) and minimal penalties. This suggests that having distinct role prompts (e.g., \emph{"Medical drone prioritizes casualties," "Logistics drone focuses on transport cost," "Infrastructure drone defends critical assets"}) can effectively divide labor when new or multiple disasters appear.
    \item \textbf{Low Diversity} yields more uniform decisions (\(\bar{d}\) is near zero), which can handle stable or low-volatility environments adequately, but fails to adapt quickly under frequent environment shifts.
\end{itemize}

Hence, moderate or strong heterogeneity among agents directly contributes to higher performance, validating that \textit{some level of viewpoint deviation is beneficial in dynamic tasks} (\textbf{answering RQ2}).

\begin{table}
\centering
\small
\caption{Scenario 1 ($N=30$): Coverage Rate (CR) for Explicit (Exp) and Implicit (Imp) Consensus across various LLM setups. Higher CR is better.}
\renewcommand{\arraystretch}{1.1}
\setlength{\tabcolsep}{5pt}
\begin{tabular}{lcc}
\toprule
\textbf{Base Model / Setup} & \textbf{S1: CR (Exp)} & \textbf{S1: CR (Imp)} \\
\midrule
Deepseek-V3         & 0.81         & \textbf{0.98}   \\
GPT-4o              & 0.79         & \textbf{0.975}  \\
GPT-4o with 4o-mini & 0.75         & \textbf{0.96}   \\
Claude-3-Sonnet     & 0.72         & \textbf{0.96}   \\
o3-mini (reasoning) & 0.70         & \textbf{0.95}   \\
GPT-4o-mini         & 0.68         & \textbf{0.95}   \\
Qwen-Plus           & 0.63         & \textbf{0.94}   \\
Llama-2             & 0.575        & \textbf{0.935}  \\
\bottomrule
\end{tabular}
\vspace{-1.5em}
\label{tab:llm_comparison}
\end{table}

\subsection{Impact of LLM Model Choice and Heterogeneous Setups}
\label{ssec:llm_impact}
Additional experiments (Scenario 1, $N=30$, Table~\ref{tab:llm_comparison}) assessed our findings across various LLMs and a mixed-model setup (GPT-4o with GPT-4o-mini). Key observations include:
\begin{itemize}[itemsep=0pt, parsep=0pt, leftmargin=*]
    \item \textbf{Consistent IC Superiority}: Implicit consensus (IC) consistently yielded higher Coverage Rates (CR) than explicit consensus (EC) across all models, including newer ones like Deepseek-V3 and the reasoning-focused o3-mini, reinforcing IC's benefit in dynamic settings.
    \item \textbf{Advanced Model Performance}: Stronger models (e.g., Deepseek-V3, GPT-4o) generally achieved higher CR under both IC and EC, but the proportional advantage of IC persisted.
    \item \textbf{Reasoning Model (o3-mini) Behavior}: The o3-mini model, designed for reasoning, performed better than GPT-4o-mini in EC (0.70 vs. 0.68) but not in IC (both 0.95). This suggests that enhanced reasoning capabilities might not always directly translate to better IC performance in dynamic agentic tasks, possibly due to \textit{overthinking}, a phenomenon noted by~\cite{Cuadron2025danger}.
    \item \textbf{Mixed-Model (GPT-4o with GPT-4o-mini)}: A hybrid team (15 GPT-4o, 15 GPT-4o-mini) performed between the homogeneous GPT-4o and GPT-4o-mini teams. In EC, its CR (0.75) was closer to the stronger GPT-4o, likely due to GPT-4o's decisive proposals influencing votes. In IC, its CR (0.96) was mid-range, reflecting an average of the mixed agents' capabilities.
\end{itemize}
These results show our core conclusions on the consensus-diversity tradeoff are robust across LLMs, with nuanced behaviors in reasoning and mixed models suggesting future research avenues.

\subsection{Generalizability Across Scenarios 2 and 3}
\label{ssec:scenarios2_3_results}
To confirm robustness, we evaluated implicit versus explicit consensus in two further dynamic scenarios: Information Spread and Manipulation (Scenario 2) and Dynamic Public Goods Provision (Scenario 3). Details are in Appendix~\ref{sec:appendix:a} and~\ref{sec:appendix:d}. Across both, implicit consensus consistently outperformed explicit consensus on key metrics. For example, in Scenario 2, it led to lower final misinformation spread and faster containment. In Scenario 3, it achieved higher public good provision rates and greater total welfare. Detailed results are in Appendix~\ref{sec:appendix:scenario2_3_results} (Tables~\ref{tab:appendix_misinfo},~\ref{tab:appendix_publicgoods}).

\subsection{Discussion and Key Insights}

\begin{enumerate} [itemsep=1pt, parsep=1pt, leftmargin=*]
    \item \textbf{Implicit vs.\ Explicit Coordination:} Figure~\ref{fig:timeseries-main} and Table~\ref{tab:consensus} confirm that \emph{implicit consensus} adapts faster to shifting disasters, achieving up to 95\% coverage in high-volatility settings.
    \item \textbf{Moderate Deviations Enhance Coverage:} Figure~\ref{fig:deviation} shows that coverage peaks at intermediate $\bar{d}$, forming a strong empirical basis for the inverted-U claim. Excessive or minimal deviations undermine synergy.
    \item \textbf{Impact of Diversity:} Medium or high diversity roles notably outperform low diversity or "no diversity," underscoring that distinct heuristics and perspectives allow drones to intercept multiple threats simultaneously rather than following a single script.
\end{enumerate}

\vspace{-0.5em}

Overall, these findings highlight the properties of dynamic consensus-diversity tradeoff. Section~\ref{subsec:deviation}'s analysis strongly supports our hypothesis that partial autonomy and role-based heterogeneity help LLM agents respond more flexibly to evolving scenarios, confirming both \textbf{RQ1} and \textbf{RQ2}. The qualitative analysis of agent dialogues in Section~\ref{ssec:dialogue_analysis_main} further illuminates how these mechanisms operate at the level of inter-agent communication and in-context reasoning, addressing \textbf{RQ3} (please refer to~\ref{ssec:dialogue_analysis_main}). The positive results across different LLM models (Section~\ref{ssec:llm_impact}) and scenarios (Section~\ref{ssec:scenarios2_3_results}) underscore the robustness of these principles.
\section{Conclusion}
We investigated \emph{when and why} \textbf{implicit consensus surpasses explicit coordination} in LLM-based MAS, emphasizing in-context learning, self-organization, and resilience with diverse LLM architectures. Our dynamic consensus-diversity model reveals that moderate, role-driven deviation from uniformity enhances performance under environmental or adversarial shifts. Experiments across dynamic disaster response, misinformation containment, and public goods provision—now including recent and reasoning-focused LLMs, plus mixed-model setups—show robust gains from emergent coordination where agents retain partial autonomy. Our work also \textbf{align with social science insights} regarding theories of limited collective common sense~\cite{whiting2024framework}. We found that complete consensus is rare and context-dependent, suggesting rigid consensus mechanisms may not work well in dynamic settings. Our experiments focused on balancing alignment with role-driven diversity. The relationship between action deviation and system performance (Section~\ref{subsec:deviation}) shows that some disagreement can be beneficial for adaptability and collective outcomes.

Future work will explore advanced dialogue strategies, formalize emergent leadership, and apply this framework to more complex, real-world MAS challenges. It is also important to investigate the theoretical underpinnings of optimal diversity thresholds in larger, more intricate agent networks, explore the dynamics of \textit{human-AI coordination} in such systems, and study the impact of adversarial agents or \textit{bad actors} within these MAS setups.

\section*{Ethical Statement}
Our scenarios involve multi-agent cooperation under dynamic conditions, including adversarial misinformation. Researchers should exercise caution when deploying such systems to ensure they do not facilitate harmful strategies (e.g., enabling misinformation). In disaster relief settings, simulated or partial deployment must account for human oversight and moral implications of decisions (e.g., triage in resource-limited contexts). This work aims to enhance collaboration mechanisms, not to displace human judgment in high-stakes scenarios.

We used ChatGPT to polish the paper. We are responsible for all the materials presented in this work.

\section*{Limitations}
(1) This work relies on purely in-context adaptation of large language models, which may struggle with extremely long dialogues or memory constraints. We also use idealized small-group scenarios, while real-world applications (e.g., large-scale social networks) may require more advanced messaging protocols. Our measure of partial diversity is approximate, and more sophisticated metrics (e.g., semantic distances in agent solutions) may yield deeper insights. Finally, controlling emergent agent behavior to ensure safety remains an open question, given the lack of a central authority in implicit consensus. (2) There are financial constraints associated with the case studies, we report the cost of a single run of these three case studies: \{\$5, \$10, \$5\}. Scalability to large agent numbers ($N \gg 100$) with current direct communication models needs further investigation, potentially requiring hierarchical or structured communication. Computational costs and API call limitations for advanced LLMs also constrained the scale and duration of some experiments. Therefore, we consider our findings as an exploratory study that needs further validation across different LLMs to enhance their generalizability.

% \section*{Acknowledgments}

% \section*{Acknowledgments}

\bibliography{anthology,custom}

\clearpage

\appendix

\section{Overall results of Scenarios 2 and 3}
\label{sec:appendix:scenario2_3_results}

Table~\ref{tab:appendix_misinfo} and Table~\ref{tab:appendix_publicgoods} present the detailed overall results for Scenario 2 (Information Spread and Manipulation) and Scenario 3 (Dynamic Public Goods Provision), respectively. These results support the generalizability of our main findings from Scenario 1, showing that implicit consensus consistently outperforms explicit consensus across various metrics in these different task domains as well.

\begin{table}
\centering
\small
\caption{Overall results of the \textbf{Information Spread and Manipulation} scenario (Scenario 2): Comparison between Explicit and Implicit Consensus. Metrics include Final Misinformation Spread Rate (MS, lower is better), Containment Time (CT, lower is better), and Coverage Diversity (CD, higher is better). Detailed discussion in Section~\ref{ssec:scenarios2_3_results} of the main paper.}
\renewcommand{\arraystretch}{1.2}
\setlength{\tabcolsep}{4pt}
\resizebox{\linewidth}{!}{
\begin{tabular}{l l ccc ccc}
\toprule
\multirow{2}{*}{\textbf{Condition}} & \multirow{2}{*}{\textbf{Level}} 
& \multicolumn{3}{c}{\textbf{Explicit Consensus}} 
& \multicolumn{3}{c}{\textbf{Implicit Consensus}} \\
\cmidrule(lr){3-5} \cmidrule(lr){6-8}
& & \textbf{MS} & \textbf{CT} & \textbf{CD} 
& \textbf{MS} & \textbf{CT} & \textbf{CD} \\
\midrule
\textbf{Overall} & - 
& 0.460 & 2.300 & 2.200 
& \textbf{0.286} & \textbf{1.411} & \textbf{2.818} \\
\midrule
\multirow{3}{*}{\textbf{Diversity}} 
& Low    
& 0.480 & 2.700 & 1.600 
& \textbf{0.322} & \textbf{1.850} & \textbf{2.000} \\
& Medium 
& 0.470 & 2.400 & 2.300 
& \textbf{0.265} & \textbf{1.367} & \textbf{2.972} \\
& High   
& 0.410 & 2.000 & 2.500 
& \textbf{0.271} & \textbf{0.986} & \textbf{2.982} \\
\midrule
\multirow{3}{*}{\textbf{Volatility}} 
& Low     
& 0.350 & 1.900 & 1.800 
& \textbf{0.300} & \textbf{1.300} & \textbf{2.010} \\
& Moderate 
& 0.480 & 2.600 & 2.500 
& \textbf{0.280} & \textbf{1.500} & \textbf{2.850} \\
& High    
& 0.550 & 2.900 & 2.700 
& \textbf{0.276} & \textbf{1.800} & \textbf{3.175} \\
\bottomrule
\end{tabular}
}
\label{tab:appendix_misinfo}
\end{table}

\begin{table}
\centering
\small
\caption{Overall results of the \textbf{Dynamic Public-Goods Provision} scenario (Scenario 3): Comparison between Explicit and Implicit Consensus. Metrics include Provision Rate (PR, higher is better), Total Welfare (TW, net total payoff, higher is better), and Free-rider Disparity (FD, lower is better). Detailed discussion in Section~\ref{ssec:scenarios2_3_results} of the main paper.}
\renewcommand{\arraystretch}{1.2}
\setlength{\tabcolsep}{4pt}
\resizebox{\linewidth}{!}{
\begin{tabular}{l l ccc ccc}
\toprule
\multirow{2}{*}{\textbf{Condition}} & \multirow{2}{*}{\textbf{Level}} 
& \multicolumn{3}{c}{\textbf{Explicit Consensus}} 
& \multicolumn{3}{c}{\textbf{Implicit Consensus}} \\
\cmidrule(lr){3-5} \cmidrule(lr){6-8}
& & \textbf{PR} & \textbf{TW} & \textbf{FD} 
& \textbf{PR} & \textbf{TW} & \textbf{FD} \\
\midrule
\textbf{Overall} & - 
& 0.765 & 21.0 & 0.190 
& \textbf{0.894} & \textbf{24.6} & \textbf{0.125} \\
\midrule
\multirow{3}{*}{\textbf{Diversity}} 
& Low    
& 0.700 & 19.2 & 0.220 
& \textbf{0.850} & \textbf{22.1} & \textbf{0.142} \\
& Medium 
& 0.770 & 21.5 & 0.180 
& \textbf{0.905} & \textbf{25.5} & \textbf{0.105} \\
& High   
& 0.810 & 23.0 & 0.160 
& \textbf{0.916} & \textbf{27.2} & \textbf{0.120} \\
\midrule
\multirow{3}{*}{\textbf{Volatility}} 
& Low     
& 0.810 & 24.0 & 0.160 
& \textbf{0.920} & \textbf{28.5} & \textbf{0.098} \\
& Moderate 
& 0.750 & 20.2 & 0.210 
& \textbf{0.898} & \textbf{24.1} & \textbf{0.135} \\
& High    
& 0.710 & 19.0 & 0.250 
& \textbf{0.870} & \textbf{22.4} & \textbf{0.150} \\
\bottomrule
\end{tabular}
}
\label{tab:appendix_publicgoods}
\end{table}

\section{Qualitative Analysis of Agent Dialogues (RQ3)}
\label{ssec:dialogue_analysis_main}
To understand \emph{how} implicit consensus fosters effective coordination and adaptation (RQ3), we analyzed the content of agent dialogues. This qualitative analysis, summarized from Appendix~D.1 of the original supplementary material, reveals key aspects of in-context learning and emergent coordination among LLM agents.

For instance, in Scenario 1 (Dynamic Disaster Response), dialogues from implicit consensus teams often showed agents explicitly referencing uncertain or contradictory reports from the environment (e.g., ``Report X says sector A is critical, but Report Y mentioned a new fire in sector B. I'll check B since others are heading to A.''). This contrasts with explicit consensus, where discussions quickly funneled towards agreement, sometimes prematurely dismissing minority viewpoints or conflicting data. Agents in implicit settings demonstrated adaptive reasoning by: 
\begin{itemize}[itemsep=0pt, parsep=0pt, leftmargin=*]
    \item \textbf{Acknowledging Uncertainty}: Agents would often voice partial information or hunches.
    \item \textbf{Distributing Cognitive Labor}: Different agents would focus on different pieces of information or potential hypotheses suggested in the discussion.
    \item \textbf{Course Correction}: An agent might initially lean towards one action but then adjust based on other agents' commitments or new information emerging in the dialogue just before action selection.
\end{itemize}
These dialogue patterns illustrate that the LLM agents' ability to process and react to nuanced linguistic information within the shared discussion is a key driver of the superior adaptability observed in the implicit consensus mode. The flexibility in interpretation and action, guided by individual roles and the collective conversation, directly contributes to better exploration and exploitation in dynamic environments, as reflected in the quantitative metrics.

\section{Detailed Experiment Scenario}
\label{sec:appendix:a}

\subsection{Dynamic Disaster Response Scenario}

\paragraph{Natural Language Information and Interference.}
Besides numeric indicators (such as severity scores), the system provides each agent a snippet of textual "reports" each round, e.g.,
\begin{quote}
\small
\emph{``Dispatch Alert: Fire intensity at Sector (3,4) may be increasing. Local residents report rising smoke. Drone \#2 previously found moderate casualties in Sector (2,2).''}
\normalsize
\end{quote}
Some reports may be incomplete or partially contradictory (e.g., a rumor that the fire is \emph{under control} despite contradictory sensor data). Agents thus need to parse these textual cues and weigh them against each other.

\paragraph{Key Experimental Factors.} \textbf{(1) Disaster severity simulation:} Each disaster has an evolving severity score $s \in [1,10]$. Higher $s$ implies higher penalty if uncontained. The environment updates $s$ in a stochastic manner, sometimes producing contradictory textual updates to test agents' ability to parse partial/misleading info. \textbf{(2) Resource constraints:} Each drone has a limited capacity (e.g., 5 units of firefighting foam). Deploying them on the wrong location wastes resources. \textbf{(3) Consensus Mechanism:} \emph{Explicit:} agents vote on one zone to be the team's priority, or follow a "unify on the most urgent location" script. \emph{Implicit:} each agent decides a location after reading the textual discussion. Some may deviate if they suspect a different site is more critical. \textbf{(4) Performance Metrics:} Coverage rate (fraction of disasters contained within 2 rounds of major severity), misallocation penalty (resources wasted on low-severity areas while ignoring high-severity ones), and average response delay.

\paragraph{Connecting to Our Research Questions.}
For Q1, we expect that under \emph{frequent} or \emph{fast-growing} disasters, implicit consensus adapts faster.  For Q2, different role prompts (e.g., "Focus on casualties" vs.\ "Minimize travel cost") introduce moderate disagreements; we measure how $\bar{d}(t)$ correlates with timely coverage. For Q3, by analyzing message logs, we see if agents revise their location choices after contradictory updates, signifying in-context learning.

\subsection{Information Spread and Manipulation Scenario}

\paragraph{Defining Misinformation.}
Misinformation is represented both as a Boolean label (node $n$ is either infected or not) and as \emph{natural language claims} that vary each round, for example:
\begin{quote}
\small
\emph{``Breaking: Node \#12 says `Vaccines have microchips', 10 neighbors are starting to share the rumor.''}
\normalsize
\end{quote}
This textual claim might be entirely false, but some "partial truths" are mixed in to raise confusion. Defender agents must interpret these claims, cross-check references, and decide which node(s) to target with a correction or "fact-check" broadcast.

\paragraph{Key Experimental Factors.}
\begin{itemize}
\item \textbf{Adversarial injections.} Every few rounds, the adversary injects new false claims into one or more nodes, sometimes disguising them as updates about a different topic.
\item \textbf{Consensus Mechanism.}  
\begin{itemize}
    \item \emph{Explicit:} defenders unify on a single node to address each round (e.g., via majority vote).
    \item \emph{Implicit:} each defender chooses a node or group of nodes to check based on discussion. Deviations can help if misinformation emerges in multiple places simultaneously.
\end{itemize}
\item \textbf{Performance Metrics.} 
\begin{itemize}
    \item \emph{Final misinformation spread} = number of nodes still misinformed after $T$ rounds.
    \item \emph{Containment time} = how many rounds it takes to isolate or correct a newly infected node.
    \item \emph{Defender coverage diversity}: how many unique nodes defenders collectively address per round.
\end{itemize}
\end{itemize}

\paragraph{Connecting to Our Research Questions.}
\begin{itemize}
    \item \textbf{Q1} Under frequent misinformation injections, forced alignment may cause defenders to chase the same node while others go unaddressed. Implicit consensus might help multi-front coverage.  
    \item \textbf{Q2} Medium or high diversity (some defenders focusing on suspicious clusters, others scanning widely) may yield better overall coverage, captured by deviation $\bar{d}(t)$.  
    \item \textbf{Q3} Round-by-round text messages allow defenders to reference past attacks ("\emph{We saw a similar rumor last round, let's watch Node 15 next}"), illustrating adaptation.
\end{itemize}

\subsection{Dynamic Public-Goods Provision Scenario}

\paragraph{Public-Good Mechanics.}
Let $x_i(t) \in [0,\,C_{\max}]$ be the amount agent $i$ contributes at round $t$, where $C_{\max}$ is the maximum individual contribution capacity. Define the \emph{total} contribution:
\[
X(t) \;=\; \sum_{i=1}^N x_i(t).
\]
A public good is considered \emph{funded} if $X(t) \geq \theta(t)$, where $\theta(t)$ is a \emph{dynamic threshold} that may change each round. When funded, the system grants a \emph{collective benefit} $B(t)$ to all agents (e.g., a large increase in safety, infrastructure, or shared profit). Each agent's net payoff from round $t$ can be expressed as:
\[
\Pi_i(t) \;=\; \underbrace{\frac{B(t)}{N}}_{\text{shared benefit}} \;-\; \underbrace{c \cdot x_i(t)}_{\text{individual cost}},
\]
where $c>0$ is the marginal cost per contribution unit (also possibly time-varying).

\paragraph{Dynamic Environment Factors.}
To incorporate volatility, we let either $\theta(t)$ (the required threshold) or $B(t)$ (the total benefit) fluctuate. For instance:
\begin{itemize}
    \item \emph{Economic Shock:} $\theta(t)$ may jump up (e.g., a crisis requiring higher funds) or drop (a technology breakthrough lowering cost).
    \item \emph{Environmental Impact:} $B(t)$ might vary based on external conditions (e.g., if the public good is a dike, storms increase the benefit of maintaining it).
    \item \emph{Rumors or Uncertain Reports:} Agents receive textual updates like "\emph{The threshold might rise to 40 next round due to a new regulation}" or "\emph{Experts claim the benefit is overestimated}," introducing partial or misleading information.
\end{itemize}

\paragraph{Consensus Mechanism.}
\begin{itemize}
    \item \textbf{Explicit Mode}: Agents vote or are instructed to adopt a single collective contribution $x_{\mathrm{group}}(t)$, which is evenly split among them. (Equivalently, they each commit to $x_i(t) = x_{\mathrm{group}}(t)/N$.)
    \item \textbf{Implicit Mode}: Agents \emph{discuss} (e.g., "I suspect we only need 20 total," "We might overshoot if the new threshold rumor is false") but finalize $x_i(t)$ independently. Some agents may deviate to free-ride or over-contribute based on their interpretation of the textual cues.
\end{itemize}

\paragraph{Performance Metrics.}
We track:
\begin{enumerate}
    \item \textbf{Provision Rate:} How often $X(t)$ meets or exceeds $\theta(t)$ across the $T$ rounds.  
    \item \textbf{Total Payoff:} $\sum_{t=1}^T \sum_{i=1}^N \Pi_i(t)$, capturing overall welfare.  
    \item \textbf{Equity or Free-Riding:} The variance or Gini coefficient of $\{x_i(t)\}$ over agents, indicating whether some consistently shoulder higher costs than others.  
\end{enumerate}
When the environment shifts threshold or benefit, a rigidly unified approach (explicit consensus) may be slow to adapt or may fail to sense incipient problems if all agents rely on the same faulty rumor. In contrast, partial diversity (some trusting a rumor, others doubting it) may maintain better long-term outcomes.

\paragraph{Connecting to Our Research Questions.}
\begin{itemize}
    \item \textbf{Q1} Under frequent or large shifts in $\theta(t)$ or $B(t)$, forced consensus might overshoot or undershoot repeatedly, while implicit consensus can allow outlier agents to either contribute more (if they believe the threshold is rising) or less (if they suspect costs are too high).
    \item \textbf{Q2} By varying agent role prompts (e.g., "always ensure public good is funded" vs.\ "minimize personal cost") we introduce moderate or strong diversity. We measure how this affects $\bar{d}(t)$ in contribution levels and see whether partial disagreement leads to more robust adaptation.
    \item \textbf{Q3} Agents may reference prior misunderstandings (\emph{"Last round we overpaid; let's not trust the rumor this time."}) or note partial contributions from others (\emph{"Agent \#2 seems to be free-riding, so I'll push my contribution up."})—clear indicators of round-by-round in-context learning.
\end{itemize}

\paragraph{Illustrative Example.}
Consider $N=5$ agents and an initial threshold $\theta(1)=30$. Four agents each propose contributing 5 units to hit 20 total, while the fifth agent, trusting a rumor that $\theta$ is lower than it looks," offers only 2. If the actual threshold is 25, then even with partial deviation, the sum (22) falls short, failing to fund the good. But if another agent, less trustful of the rumor, deviates upward to 7 units, the total might reach 24—still not enough. Over time, the group's discussion leads them to converge around 25 or more, but occasionally someone might keep free-riding. Meanwhile, a random shock might raise $\theta(5)$ to 40; if all adopt the same unchanging strategy, the good is unfunded. If one agent is "paranoid," contributing extra, it may save the collective from shortfall. This scenario thus highlights how partial autonomy can hedge against rumor-driven errors or incomplete knowledge.

\section{Experimental Settings}
\label{sec:appendix:b}

This appendix provides the concrete experimental configurations for our three case studies, including environment parameters, reward/penalty functions, and example prompts. Unless noted otherwise, each experiment is repeated over 5 random seeds (or distinct initializations) to reduce variance, and results are averaged. For all scenarios and all runs, the model parameters \texttt{temperature} is set to \texttt{0.7} to balance the performance and diversity, and the \texttt{max\_token} is set to \texttt{256}.

\subsection{Dynamic Disaster Response}
\label{appendix:disaster}

\paragraph{Grid and Disaster Zones.}
We use a $10\times10$ grid representing a simplified city map. At any point, there are up to $K=3$ active disasters (e.g., fires, floods). The environment updates \emph{every round} by:
\begin{itemize}
    \item Potentially moving an existing disaster to a neighboring grid cell (random direction).
    \item Changing the \emph{severity} $s \in [1,10]$ of one or more disasters (can increase or decrease by 1--3 points).
    \item Creating a new disaster with small probability $p_{\text{new}}=0.2$ if fewer than 3 are active.
\end{itemize}
Each disaster occupies a single cell, but severity influences how damaging it is if not contained.

\paragraph{Agent Roles and Prompts.}
We have $N=3$--$100$ LLM agents (GPT-4, Claude, Llama-2, Qwen) controlling "drones." Each agent is given a short role prompt, such as:
\begin{itemize}
    \item \texttt{Medical drone:} \emph{``Focus on rescuing casualties in the highest-severity disaster zone for people.''}
    \item \texttt{Infrastructure drone:} \emph{``Protect power lines and roads. Even if severity is high elsewhere, prioritize built structures.''}
    \item \texttt{Logistics drone:} \emph{``Minimize travel cost. Quickly move to the nearest active zone if severity is above 5.''}
\end{itemize}
In the \textbf{low-diversity} condition, all agents share a nearly identical prompt (e.g., "Always address the highest severity zone"). In \textbf{medium-diversity}, two or three distinct prompts exist. In \textbf{high-diversity}, each agent has a unique role with potentially conflicting heuristics.

\paragraph{Communication and Textual Interference.}
Each round, a textual "situation report" is provided, e.g.:
\begin{quote}
\small
\emph{``A large fire at (3,4) has severity 8. Some witnesses claim the fire is spreading north. Another source says no sign of growth. Casualties reported near (3,5).''}
\normalsize
\end{quote}
Up to 20\% of these messages may be \emph{contradictory or incomplete}. Agents must interpret them carefully. In explicit consensus mode, a final "team vote" or forced alignment prompt merges all votes into a single chosen cell. In implicit mode, each drone chooses a cell independently.

\paragraph{Rewards and Penalties.}
\begin{itemize}
    \item \textbf{Disaster Containment:} If a drone visits the grid cell of a disaster of severity $s$ and stays there for 1 full round, the severity of that disaster is reduced by up to 3 points. Once $s\le0$, the disaster is "cleared," yielding $+s \times \alpha$ (e.g., $\alpha=5$) as a reward. (This is a positive number since $s$ was originally $>0$.)
    \item \textbf{Uncontained Penalty:} Each round a severity-$s$ disaster remains active, it incurs a penalty $-s \times \beta$ (e.g., $\beta=2$).
    \item \textbf{Misallocation Cost:} If more than $M=2$ drones converge on the same location while another active disaster is \emph{uncovered}, a penalty $-5$ is applied that round (representing wasted resources).
\end{itemize}
We log the \emph{overall net reward} (total containment benefits minus penalties) after $T=20$ or $30$ rounds, as well as \emph{time-series} data on which cells each drone chose (to compute $d_i(t)$).

\paragraph{Volatility Settings.}
\begin{itemize} [itemsep=1pt, parsep=1pt, leftmargin=*]
    \item \textbf{Low Volatility}: Disasters rarely move (once every 3 rounds), severity changes are small ($\pm1$). 
    \item \textbf{Moderate Volatility}: Disasters can move or spawn every 2 rounds; severity can jump by up to $\pm2$.
    \item \textbf{High Volatility}: Every round sees at least one shift or new disaster with severity changes up to $\pm3$.
\end{itemize}
We expect implicit consensus to shine in moderate/high volatility, where strictly unifying on a single plan may lead to slow adaptation or over-allocation to one zone.

\subsection{Information Spread and Manipulation}
\label{appendix:infospread}

\paragraph{Network and Misinformation Mechanics.}
We generate a \textbf{scale-free} network with $50$ nodes. Each node can be in state \{\emph{unaware}, \emph{informed}, \emph{misinformed}\}. Initially, 2--5 random nodes are "misinformed" by the adversarial agent. After each round:
\begin{enumerate} [itemsep=1pt, parsep=1pt, leftmargin=*]
    \item Each misinformed node may infect its neighbors with probability $p_{\text{spread}} = 0.2$ unless a neighbor has been "fact-checked" this round.
    \item The adversarial agent may inject a new rumor into $k_{\text{new}}=1$--2 additional nodes, typically accompanied by a textual snippet (e.g., \emph{``Secret leak: Node \#20 claims vaccines contain microchips''}).
\end{enumerate}
We continue for $T=20$ rounds or until $>80\%$ of nodes are infected (which terminates the simulation if defenders fail).

\paragraph{Defender Agents.}
We have $N=3$--$10$ defender LLM agents, each controlling a "monitoring bot." Every round, each agent can:
\[
a_i(t) \;=\; \{\text{choose up to } R \text{ nodes to fact-check}\},
\]
with $R=3$ by default. In \textbf{explicit} mode, the defenders unify on a single set of nodes (e.g., via majority vote on which $R$ nodes to check). In \textbf{implicit} mode, each agent decides individually but may coordinate through textual discussion. Agents see partial updates like:
\begin{quote}
\small
\emph{``Suspicious rumor detected at Node \#15. A new wave of misinformation might have reached Node \#29. Some people say Node \#29 was already vaccinated with correct info.''}
\normalsize
\end{quote}
Some updates are contradictory or ambiguous, fostering potential disagreement on which nodes are truly at risk.

\paragraph{Performance Metrics.}
\begin{itemize}
    \item \textbf{Final Misinformation Spread}: Percentage of nodes misinformed at the end of $T$ rounds.
    \item \textbf{Containment Time}: The average number of rounds needed to reduce an outbreak from $m$ newly infected nodes to $<m/2$.
    \item \textbf{Coverage Diversity}: At each round, how many \emph{unique} nodes got fact-checked across all defenders (higher is typically better if multiple rumors exist).
\end{itemize}

\paragraph{Diversity Conditions.}
\begin{itemize} [itemsep=1pt, parsep=1pt, leftmargin=*]
    \item \textbf{Low diversity}: All defenders share the same heuristic (e.g., "prioritize highest-degree suspicious node"), making them converge easily.
    \item \textbf{Medium diversity}: Some defenders do broad scanning, others do targeted local checks. 
    \item \textbf{High diversity}: One or two defenders might have contradictory priorities (e.g., "flag \emph{any} node that had rumors last round," ignoring new ones). 
\end{itemize}
We analyze how these differences lead to partial deviation ($\bar{d}(t)$) in which nodes are tackled each round, and whether that boosts or impairs containment.

\paragraph{Volatility Settings.}
\begin{itemize}
    \item \textbf{Low}: Adversary injects new misinformation only every 4 rounds; $p_{\text{spread}}=0.1$.
    \item \textbf{Moderate}: Injections every 2--3 rounds; $p_{\text{spread}}=0.2$.
    \item \textbf{High}: Injections nearly every round; $p_{\text{spread}}=0.3$.
\end{itemize}

\subsection{Dynamic Public-Goods Provision}
\label{appendix:publicgood}

\paragraph{Basic Setup.}
We have $N=3$--$30$ LLM agents that each round decide an investment $x_i(t) \in [0, C_{\max}]$. If the total $X(t) = \sum_{i=1}^N x_i(t)$ meets or exceeds a threshold $\theta(t)$, a public good is "funded," yielding a benefit $B(t)$ shared among agents.

\paragraph{Cost and Benefit Functions.}
\begin{itemize}
    \item \textbf{Threshold} $\theta(t)$: starts at $\theta(1)=30$ (for $N=5$) and may shift by $\pm 5$ or $\pm 10$ at random intervals to simulate external events (e.g., new government regulations). 
    \item \textbf{Benefit} $B(t)$: typically 100 if funded, else 0. We sometimes allow $B(t)$ to fluctuate between 80 and 120 to represent environmental or economic factors.
    \item \textbf{Individual payoff} for agent $i$ at round $t$:
    \[
    \Pi_i(t) = \begin{cases}
    \tfrac{B(t)}{N} - c\, x_i(t), & \text{if } X(t) \ge \theta(t), \\
    -c\,x_i(t), & \text{otherwise},
    \end{cases}
    \]
    with $c=1$ or $2$ for cost per unit contribution. 
\end{itemize}

\paragraph{Communication and Textual Uncertainty.}
Before choosing $x_i(t)$, each agent receives ambiguous or noisy reports about $\theta(t)$ or $B(t)$:
\begin{quote}
\small
\emph{``Analyst warns threshold could jump to 40 next round. Another says \emph{`No, it remains 30'}.''}
\normalsize
\end{quote}
In \textbf{explicit} mode, the group merges all votes into a single contribution value $x_{\mathrm{group}}(t)$, which each agent pays evenly. In \textbf{implicit} mode, each agent decides $x_i(t)$ independently after reading the discussion. Some might deviate to "cover the gap" if they suspect others will under-contribute.

\paragraph{Performance Metrics.}
\begin{itemize}
    \item \textbf{Provision Rate}: fraction of rounds where $X(t)\ge\theta(t)$.
    \item \textbf{Total Welfare}: $\sum_{t=1}^T \sum_{i=1}^N \Pi_i(t)$.
    \item \textbf{Contribution Distribution}: standard deviation or Gini of $\{x_i(t)\}$, revealing potential free-riding.
\end{itemize}

\paragraph{Diversity Conditions.}
\begin{itemize}
    \item \textbf{Low}: All agents have near-identical role prompts (``aim to exactly meet the threshold''). 
    \item \textbf{Medium}: Some are more risk-averse (contribute extra), others more cost-sensitive. 
    \item \textbf{High}: Strongly conflicting roles (``always contribute minimal'' vs.\ ``guarantee coverage by overshooting''), plus a "moderate" agent.
\end{itemize}
We compare \emph{implicit} vs.\ \emph{explicit} across different thresholds' volatility to see if partial deviation yields more stable funding despite uncertain information.

\subsection{Common Protocol and Logging}
For each scenario, we run $T=20$ or $T=30$ rounds:
\begin{enumerate}
    \item \textbf{Environment Update}: The simulation changes state (disaster severity, misinformation injections, or threshold shifts).
    \item \textbf{Report Generation}: A textual summary (and possibly contradictory rumors) is sent to each agent.
    \item \textbf{Discussion Phase}: Agents produce up to $K$ messages each (where $K\in\{1,2\}$ typically), referencing the new info and proposing strategies.
    \item \textbf{Action Phase}: In \textbf{explicit} consensus, a final vote or forced agreement yields one uniform action or plan. In \textbf{implicit}, each agent decides its own $a_i(t)$ after reading the messages.
    \item \textbf{Reward/Penalty Computation}: We apply the scenario-specific reward/penalty rules (Sections~\ref{appendix:disaster}-\ref{appendix:publicgood}) and log:
    \begin{itemize}
        \item Agent actions $\{a_i(t)\}$, used to compute $\bar{d}(t) = \frac{1}{N}\sum_i\|a_i(t) - \mu(t)\|$.
        \item Scenario performance metrics (coverage, spread, or public-good provisioning).
    \end{itemize}
\end{enumerate}
Each experimental condition (low/medium/high diversity, low/medium/high volatility, implicit/explicit mode, etc.) is repeated over multiple random seeds. We collate the final performance averages and produce \emph{time-series} plots of $(\bar{d}(t), \text{performance}(t))$.

\subsection{Sample Prompts and Roles}
Below is an illustrative snippet of role prompts for one scenario. The actual implementation uses variants for each condition.

\begin{description}
\item[Medical Drone Prompt (Disaster):] 
\emph{``You are a Medical Drone focused on saving human lives. You have limited medical kits. Always prioritize zones with potential casualties. If multiple high-severity disasters exist, choose the one with the greatest threat to people.''}

\item[Infrastructure Drone Prompt (Disaster):]
\emph{``You are an Infrastructure Drone. Your mission is to prevent damage to critical facilities (power grid, roads). Even if the severity is high elsewhere, you prefer protecting large-scale infrastructure for the long run.''}
\end{description}

\noindent Similar or contrasting prompts are used to induce different priorities and cause moderate or high disagreement within the group.

\subsection{Implementation Details}
We use a custom Python environment for each scenario, with round-by-round updates. Agents interface via API calls to LLMs (GPT-4, Claude, Llama-2, Qwen). Each agent's messages are truncated or summarized to maintain manageable context length. No fine-tuning or parameter training is performed; all adaptation emerges purely through repeated textual interactions (i.e., in-context learning). Further low-level details (including random seeds, exact parameter tables, and examples of message transcripts) will be released as supplemental material.

\subsection{Evaluation Methodology}
After each run:
\begin{itemize}
    \item We compute aggregated performance metrics (net reward, final spread, total payoff) to compare \textbf{implicit} vs.\ \textbf{explicit} consensus. 
    \item We examine how agent-level $\bar{d}(t)$ evolves. A typical analysis might cluster rounds based on environment shocks (e.g., times when a new disaster spawns or threshold jumps), to see how quickly the system re-stabilizes.
    \item We optionally analyze \emph{dialogue transcripts} for qualitative insights on how agents reference prior mistakes or respond to contradictory info (testing \textbf{Q3} regarding in-context adaptation).
\end{itemize}
These combined quantitative and qualitative measures allow us to test the dynamic consensus-diversity tradeoff hypotheses described in the main paper.

\section{Simplified Theoretical Model of Consensus-Diversity Dynamics}
\label{app:theoretical_model}

In this appendix, we present a minimal random-iteration model for studying the consensus--diversity tradeoff in a more analytically tractable setting. While the main paper's results focus on \textbf{LLM-driven multi-agent systems} (where agent ``diversity'' arises from distinct roles and textual reasoning), this simplified model provides insight into the effect of purely random deviations on consensus formation.

\paragraph{Motivation and Precedents.}
Classical multi-agent consensus models~\cite{degroot1974reaching,olfati2007consensus} typically assume each agent updates its state by averaging neighbors' values. However, in highly dynamic or uncertain environments, agents may also exhibit random drifts or maintain individual preferences (``stubbornness''). Inspired by related stochastic models in opinion dynamics~\cite{friedkin2011social, hegselmann2002opinion}, we introduce:
\[
    x_i(t+1) 
    \;=\; 
    (1-\alpha)\,x_i(t) 
    \;+\;
\]
\[
    \alpha \,\mu(t)
    \;+\;
    \gamma\,[\,a^*(t)-x_i(t)\,]
    \;+\;
    \beta \,\epsilon_i(t),
\]
where:
\begin{itemize}[itemsep=0pt, parsep=0pt, leftmargin=*]
    \item $x_i(t)$ is agent $i$'s scalar state (or opinion) at time $t$,
    \item $\mu(t)=\tfrac{1}{N}\sum_{j=1}^N x_j(t)$ is the group mean,
    \item $\alpha \in [0,\,1]$ is a consensus weight pulling each $x_i(t)$ toward $\mu(t)$,
    \item $\gamma \geq 0$ is a \emph{pull strength} toward the environment's current optimum $a^*(t)$. We set $\gamma \geq 0$ to ensure the group not only tends toward an internal consensus but also tracks the external environment optimum $a^*(t)$. This modification better simulates the scenario where agents receive some feedback about the correct direction, allowing for a potential "optimal" level of exploration $\beta$ that balances quick convergence and adaptability,
    \item $\beta \geq 0$ scales the random "diversity" or noise term $\epsilon_i(t)\sim \mathcal{N}(0,1)$.
\end{itemize}
This iteration is a toy abstraction for "consensus plus partial diversity," omitting the richer \emph{semantic} differences that LLM agents exhibit in the main text. Nevertheless, it allows us to explore how random deviations interact with a basic alignment mechanism.

\begin{figure}
    \centering
    \includegraphics[width=1\linewidth]{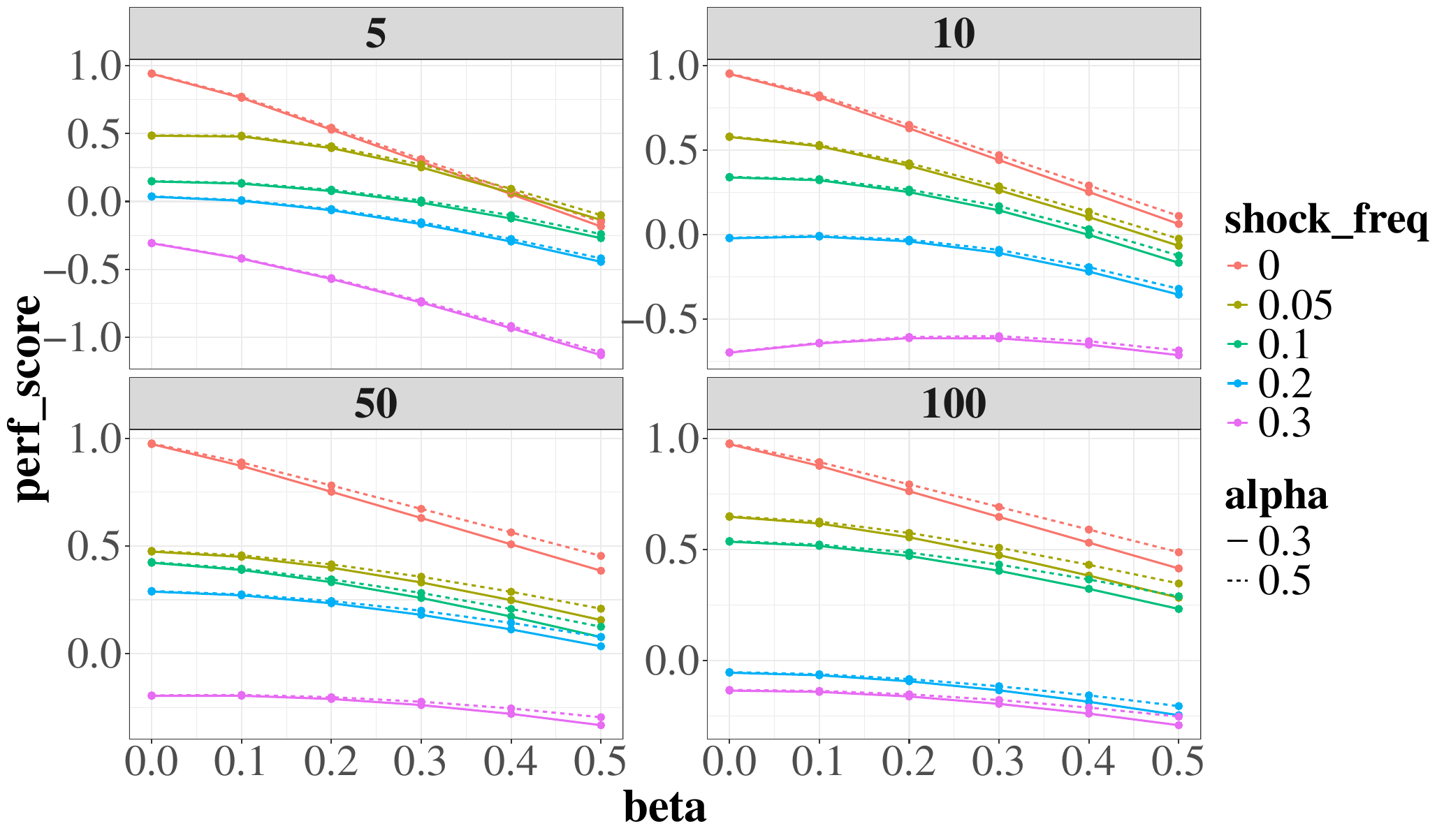}
    \vspace{-1.5em}
    \caption{Simplified Theoretical Model: Performance vs. Beta by Agent Count (N) Under Various Shock\_freq \& Alpha.}
    \vspace{-1.5em}
    \label{fig:toy-model-example}
\end{figure}

\paragraph{Environment Shocks.}
To simulate a \emph{dynamic} optimum $a^*(t)$, we let it evolve according to random shocks:
\begin{equation}
\text{if}~\mathrm{rand}() < \mathrm{shock\_freq}, \quad a^*(t+1) \leftarrow a^*(t) + \Delta,
\end{equation}
where $\Delta$ is sampled uniformly in some interval (\emph{e.g.}, $[-1,\,1]$). One may also keep $a^*(t)$ fixed when no shock occurs. This mimics an external environment whose "best action" changes unpredictably.

\paragraph{Performance Metrics.}
After $T$ rounds, we measure:
\begin{itemize}[itemsep=0pt, parsep=0pt, leftmargin=*]
    \item \textbf{Average distance to optimum}:     
        $\displaystyle \overline{D_{\text{opt}}} = \frac{1}{T}\sum_{t=1}^T \tfrac{1}{N}\sum_{i=1}^N \bigl|\,x_i(t) - a^*(t)\bigr|$.
    A smaller value indicates better tracking of the environment optimum.
    \item \textbf{Average agent deviation}:
    $\displaystyle \overline{d} = \frac{1}{T}\sum_{t=1}^T \tfrac{1}{N}\sum_{i=1}^N \bigl|\,x_i(t)-\mu(t)\bigr|$.
    This indicates how divergent agents remain from the group mean.
    \item \textbf{Simple performance score}:
    $\displaystyle \text{perf\_score} = 1 - \overline{D_{\text{opt}}}~,$
    which can be negative if $\overline{D_{\text{opt}}}>1$.
\end{itemize}

\paragraph{Empirical Trends and Limitations.}
Figure~\ref{fig:toy-model-example} (in the main paper) shows typical outcomes of sweeping $(\alpha,\beta,\mathrm{shock\_freq},\gamma)$:
\begin{itemize} [itemsep=0pt, parsep=0pt, leftmargin=*]
    \item When $\beta=0$ (\emph{no random deviation}), the group quickly converges to a single value; if $\gamma>0$ and shocks are mild, they track $a^*(t)$ well. In stable environments, this yields high performance.
    \item As $\beta$ grows, purely random noise tends to \emph{worsen} performance, especially if $\gamma$ is small, because the group states drift around widely, failing to coalesce near $a^*(t)$.
    \item Even under moderate shocks, we do not observe an "inverted-U" benefit purely from random $\beta$; performance typically declines monotonically in $\beta$. 
\end{itemize}
This stands in contrast to the main paper's \emph{dynamic consensus-diversity tradeoff}, where partial diversity arises from \emph{cognitively informed} dissent rather than uniform Gaussian noise. In other words, this random-iteration model confirms that \emph{unstructured} or \emph{white-noise} deviations generally hurt consensus. By itself, it \textbf{does not} show the potential \emph{benefits} of moderate disagreement or role-specific exploration.

We note that when the environment feedback term $\gamma$ is sufficiently large relative to the shock amplitude, a small to moderate amount of noise ($\beta$) can actually \emph{improve} adaptation, rather than degrade it. In other words, there exists a narrow parameter region where partial diversity from random perturbations helps the group track the shifting optimum $a^*(t)$ more effectively. Although this phenomenon remains less pronounced than in our LLM-driven multi-agent experiments (where diversity is semantically grounded), it does illustrate that an "optimal" $\beta$ can arise even in a purely random-iteration model, provided $\gamma$ and the shock parameters are appropriately balanced.

\paragraph{Interpretation for Our Work.}
In the main experiments (Sections~4 and 5 of the main paper), we highlight that \emph{LLM-based} agents with distinct roles produce \emph{meaningful disagreements}, which can facilitate adaptability in shifting environments. The toy iteration model here clarifies that \emph{if} diversity is solely random, performance typically decreases with $\beta$. Thus:
\begin{enumerate}[itemsep=0pt, parsep=0pt, leftmargin=*]
    \item \textbf{No contradiction:} The simplified model's \emph{monotonic} decline underscores how random "noise" alone undermines stable consensus, especially under frequent shocks.
    \item \textbf{Need for structured diversity:} Real LLM agents do not \emph{merely} add random perturbations; they incorporate textual cues, role differences, and partial knowledge--often generating beneficial exploration. 
\end{enumerate}
Therefore, while the random-iteration model is convenient for partial theoretical analysis (one can show conditions for convergence in expectation when $\beta$ is small, etc.), it does \emph{not} replicate the emergent synergy from purposeful agent heterogeneity. Future extensions could incorporate strategic exploration or role-based logic in a more advanced "consensus + diversity" iteration model, potentially revealing the inverted-U phenomenon seen in structured multi-agent dialogues.

\paragraph{References for Dynamic Iteration Models.}
The approach here is loosely related to classic work on \textbf{DeGroot} averaging~\cite{degroot1974reaching} and \textbf{Friedkin--Johnsen} "stubbornness" models~\cite{friedkin2011social}, extended to include environment shifts and additive noise. For comprehensive surveys on consensus protocols and opinion dynamics, see~\cite{olfati2007consensus, hegselmann2002opinion}.

% End of new Appendix C (Simplified Theoretical Model).
% User should check the rest of their appendix file for old sections D, E, F.
% Section D (Prompt Example and Dialogue Analysis) might be kept if it has more detail.
% Sections E (Old Scenario 2/3 results) and F (Old LLM table) should be removed.

% ... (The rest of the appendix file, if any) ...

\section{Performance of Different Base models}
\label{sec:appendix:f}

In summary, from Table~\ref{tab:basemodels} we can figure out that while \texttt{GPT-4o} consistently leads in overall performance, every base model achieves better results under implicit consensus, reaffirming the advantage of partial autonomy across all three scenarios.

\begin{table}[h]
\centering
\small
\caption{Performance of Different Base Models across the three scenarios. 
S1 (CR) is Coverage Rate,
S2 (MS) is Misinformation Spread,
S3 (PR) is Provision Rate.
``Exp'' and ``Imp'' refer to explicit vs.\ implicit consensus.}
\renewcommand{\arraystretch}{1.15}
\setlength{\tabcolsep}{5pt}
\resizebox{\linewidth}{!}{%
\begin{tabular}{l cc cc cc}
\toprule
\multirow{2}{*}{\textbf{Base Model}} 
& \multicolumn{2}{c}{\textbf{S1: CR}} 
& \multicolumn{2}{c}{\textbf{S2: MS}} 
& \multicolumn{2}{c}{\textbf{S3: PR}} \\
\cmidrule(lr){2-3}\cmidrule(lr){4-5}\cmidrule(lr){6-7}
& \textbf{Exp} & \textbf{Imp}
& \textbf{Exp} & \textbf{Imp}
& \textbf{Exp} & \textbf{Imp} \\
\midrule
\textbf{GPT-4o} 
 & 0.79  & 0.975 
 & 0.40  & 0.20 
 & 0.82  & 0.92 \\
\textbf{Claude-3-Sonnet} 
 & 0.72  & 0.96 
 & 0.43  & 0.25 
 & 0.79  & 0.91 \\
\textbf{GPT-4o-mini} 
 & 0.68  & 0.95 
 & 0.46  & 0.27 
 & 0.76  & 0.885 \\
\textbf{Qwen-Plus} 
 & 0.63  & 0.94 
 & 0.49  & 0.33 
 & 0.74  & 0.875 \\
\textbf{Llama-2} 
 & 0.575 & 0.935
 & 0.52  & 0.38 
 & 0.715 & 0.88 \\
\midrule
\textbf{Average}
 & 0.679 & 0.952
 & 0.46  & 0.286
 & 0.765 & 0.894 \\
\bottomrule
\end{tabular}%
}
\label{tab:basemodels}
\end{table}

\section{Prompt Example and Dialogue Analysis}
\label{sec:appendix:d}

\subsection{Dialogue Analysis on Dynamic Disaster Response (RQ3)}
\label{sec:analysis-rq3}

To address \textbf{RQ3}---how agents coordinate and revise their decisions in context based on each other's statements---we examined select rounds from the agent interaction log. Below, we highlight three observations demonstrating that \emph{partial disagreement} and role-driven perspectives lead to adaptive, cooperative behavior.

\paragraph{(1) Role-Specific Choices Lead to Divergent Actions but Rapid Coverage.}
In the very first round (\texttt{round=0}), each agent independently selects different grid coordinates:
\begin{itemize} [itemsep=1pt, parsep=1pt, leftmargin=*]
    \item \textbf{Drone 0 (Medical)} moves to \texttt{[5,5]} to address ``immediate casualty evacuation''.
    \item \textbf{Drone 1 (Infrastructure)} chooses \texttt{[6,6]} to ``ensure power lines and roads remain functional''.
    \item \textbf{Drone 2 (Logistics)} goes to \texttt{[4,5]} for ``comprehensive coverage around high-severity zones''.
\end{itemize}
These decisions show that even in the same round, they do not unify on a single location but rather \emph{diverge} based on role priorities. As a result, multiple key zones are covered simultaneously. 

\paragraph{(2) Agents Adapt Their Plans After Reading Others' Messages.}
By \texttt{round=2}, the Medical drone has chosen \texttt{[5,3]}, while Infrastructure and Logistics drones pick \texttt{[6,5]} or \texttt{[6,5]} respectively. Examining their messages, we find explicit references to each other's stated actions:
\begin{quote}
\emph{Drone~2 (Logistics)}: ``I am positioning at [6,5] to assist with infrastructure preservation, ensuring we prevent potential overlap...''
\end{quote}
This highlights how reading other drones' allocations (e.g., "someone is already at the casualty zone") motivates partial shifts in coverage. Rather than forcing a single group plan, the system allows each drone to deviate if it sees unaddressed needs elsewhere.

\paragraph{(3) Ongoing Coordination Prevents Over-Concentration.}
At later rounds (e.g., \texttt{round=7} and \texttt{round=8}), the Medical drone announces high-severity casualties in zones like \texttt{[7,2]} or \texttt{[7,8]}, while Infrastructure and Logistics drones opt for \texttt{[5,3]} or \texttt{[5,6]} to handle different concerns. Their messages show active avoidance of unnecessary overlap:
\begin{quote}
\emph{Drone~1 (Infrastructure)}: ``I will focus on securing power lines and roads near [5,3]... Continue providing support at [7,2].''
\end{quote}

Thus, partial disagreement again drives \emph{diverse coverage}, ensuring that each critical need (casualty rescue, infrastructure defense, supply logistics) is handled. The drones repeatedly \emph{reference} one another's chosen actions to avoid duplication, demonstrating a form of emergent in-context negotiation.

\paragraph{Summary for RQ3.}
These logs confirm that (a) each agent's specialized role leads to distinct decisions, (b) the presence of partial disagreement triggers broader coverage of dynamic hazards, and (c) agents revise their actions in response to dialogue updates rather than following a single script. Consequently, the system remains flexible, distributing resources where they are most needed while avoiding premature consensus on one zone. This supports our claim that implicit consensus structure and in-context learning ability of LLM agents fosters robustness through continuous adaptation and partial autonomy among agents.

\subsection{Case Study 1: Dynamic Disaster Response}
\begin{dialogue}{Prompt Template}
You are Drone \{id\}, a \{role\} in a disaster re\-sponse team.

Current situation: \{grid description and dis\-aster states\}

Other drone messages: \{messages from other drones\}

Your role instructions: \{role-specific guide\-lines\}

Based on the current situation and your role, provide:\\
1. Your analysis of the situation\\
2. Your proposed action as grid coordinates [x,y]\\
3. A brief message to share with other drones

Format your response as JSON exactly like this example:\\
\{  "analysis": "My analysis of the situation...",\\
    "action": [3,4],\\
    "message": "My message to other drones..."\}
\end{dialogue}

\begin{dialogue}{Role Types}
\textbf{Medical Drone:}\\
"Focus on rescuing casualties in highest-severity disaster zones for people."

\textbf{Infrastructure Drone:}\\
"Protect power lines and roads. Even if severity is high elsewhere, prioritize built structures."

\textbf{Logistics Drone:}\\
"Minimize travel cost. Quickly move to nearest active zone if severity is above 5."
\end{dialogue}

\begin{dialogue}{Agent Actions}
$\bullet$ Choose grid coordinates [x,y] to move to\\
$\bullet$ Analyze situation severity\\
$\bullet$ Share tactical information with other drones
\end{dialogue}

\subsection{Case Study 2: Information Spread and Manipulation}
\begin{dialogue}{Prompt Template}
You are Defender \{id\}, a \{role\} in an information manipulation defense team.

Current situation: \{network state and spread description\}

Network information: \{structure and metrics\}

Other defender messages: \{messages from other defenders\}

Based on the situation and your role, provide:\\
1. Your analysis of the network state\\
2. Your proposed nodes to fact-check [maximum 3]\\
3. A brief message to share with other defenders

Format your response as JSON exactly like this example:\\
\{  "analysis": "My analysis of the situation...",\\
    "target\_nodes": [1, 4, 7],\\
    "message": "My message to other defenders..." \}
\end{dialogue}

\begin{dialogue}{Role Types}
\textbf{Proactive Defender:}\\
"Prioritize checking high-influence nodes before they get infected. Focus on creating network firebreaks."

\textbf{Reactive Defender:}\\
"Target nodes that are actively spreading misinformation. Focus on reducing current spread."

\textbf{Network Analyzer:}\\
"Study network structure and identify critical nodes. Track infection patterns."

\textbf{Rapid Responder:}\\
"Quickly respond to new infections. Focus on containing new outbreaks."
\end{dialogue}

\begin{dialogue}{Agent Actions}
$\bullet$ Select nodes to fact-check (maximum 3 per round)\\
$\bullet$ Analyze spread patterns\\
$\bullet$ Share strategic insights about network vulnerabilities
\end{dialogue}

\subsection{Case Study 3: Dynamic Public-Goods Provision}
\begin{dialogue}{Prompt Template}
You are Contributor \{id\}, a \{role\} in a public goods provision team.

Current situation: \{threshold and benefit description\}

Previous outcomes: \{last round results\}

Other contributor messages: \{messages from other contributors\}

Based on the situation and your role, provide:\\
1. Your analysis of the situation\\
2. Your proposed contribution amount [0-\{max\_contribution\}]\\
3. A brief message to share with other contributors

Format your response as JSON exactly like this example:\\
\{"analysis": "My analysis of the situation...",\\
    "contribution": 10.5,\\
    "message": "My message to other contributors..."\}
\end{dialogue}

\begin{dialogue}{Role Types}
\textbf{Altruistic:}\\
"Prioritize meeting the threshold to ensure public good provision. Willing to contribute more than fair share."

\textbf{Strategic:}\\
"Balance personal costs against public benefits. Adjust contributions based on others' behavior."

\textbf{Conservative:}\\
"Prefer smaller, safer contributions. Focus on sustainable long-term participation."

\textbf{Adaptive:}\\
"Quickly adjust to threshold and benefit changes. Learn from past outcomes."
\end{dialogue}

\begin{dialogue}{Agent Actions}
$\bullet$ Decide contribution amount [0, max\_contribution]\\
$\bullet$ Analyze group dynamics\\
$\bullet$ Share strategic insights about optimal contribution levels
\end{dialogue}

\end{document}